\documentclass[hyper,a4paper]{JHEP3}
\usepackage{graphicx}
\usepackage{graphics}
\usepackage[mathscr]{eucal}
\usepackage{amssymb}
\usepackage{amsmath}
\usepackage{xspace}
\usepackage{listings}
\usepackage{ulem}
\usepackage{slashed}
\usepackage{verbatim}
\usepackage{axodraw4j}
\usepackage{caption}
\usepackage{subcaption}
\usepackage{multirow}
\title{Resonant mono Higgs at the LHC}

\author{
Lorenzo Basso$^{a\ast}$\\
(a) Universit\`e de Strasbourg, IPHC, 23 rue du Loess 67037 Strasbourg,
      France \\
Institut Pluridisciplinaire Hubert Curien/D\'epartement Recherches 
Subatomiques,
CNRS-IN2P3, UMR 7178, 23 Rue du Loess, F-67037 Strasbourg, France \\
CPPM, Aix-Marseille Universit\'e, CNRS-IN2P3, UMR 7346, 163 avenue de Luminy, 13288 Marseille Cedex 9, France]

$\ast$Email: \email{basso@cppm.in2p3.fr} \\
}
\abstract{In recent years, the production of a SM particle with large missing transverse
momentum, dubbed mono-X searches, have gained increasing attention. After the discovery
of the Higgs boson in 2012, the run-II of the LHC will now scrutinise its properties, looking
for BSM physics. In particular, one could search for mono-Higgs signals, that are typically
studied in models addressing dark matter. However, this signal can appear also in models
addressing the neutrino masses, if additional heavier neutrinos with masses at the
electroweak scale are present. The latter will couple to
the SM neutrinos and the Higgs boson, yielding a type of mono-Higgs signal not considered
for dark matter: the resonant production of a Higgs boson and missing energy. In this paper, we
address the LHC exclusion power of the latter with dedicated detector simulations, and
reinterpret it in a benchmark scenario for neutrino mass generation.}

\keywords{Higgs, BSM, New Physics, Neutrinos}

\begin{document}
%\maketitle
%\tableofcontents
% ==============================================================================

\section{Introduction}
Searches of single objects produced in association with missing energy, often dubbed {\it mono-$X$}, have gained considerable amount of attention in recent years as powerful probes for New Physics. Most notably, $X$ can be a standard model (SM) particle, such as
a light jet \cite{Beltran:2010ww,Goodman:2010ku,Rajaraman:2011wf,Fox:2011pm,Khachatryan:2014rra,Aad:2015zva,Abdallah:2015uba},  
a $b$-quark \cite{Lin:2013sca,Aad:2014vea}, 
a top-quark \cite{Andrea:2011ws,Agram:2013wda,Boucheneb:2014wza},
a $W$ boson \cite{Bai:2012xg,Aad:2013oja,ATLAS:2014wra,Khachatryan:2014tva},
a $Z$ boson \cite{Bell:2012rg,Carpenter:2012rg,Aad:2014vka,Alves:2015dya,No:2015xqa,Abdallah:2015uba},
a photon ($\gamma$) \cite{Fox:2011pm,Khachatryan:2014rwa,Aad:2014tda,Abdallah:2015uba},
or a Higgs boson~\cite{Carpenter:2013xra,Petrov:2013nia,Berlin:2014cfa,No:2015xqa}.

These mono-$X$ signatures are mostly inspired by dark matter (DM) models, where the
additional source of missing energy is provided by the DM candidates that are produced in
association with it. The various signatures are described by means of non-renormalisable
operators or simplified/effective models. In this paper we focus on the latter signature, the
mono-Higgs case. With the recent discovery of the Higgs boson, a large effort is going to
be devoted in the run-II of the LHC to scrutinise its properties in great details. Furthermore,
any new particle that has mass should in some way couple to the Higgs boson. In this sense, the
Higgs boson is a generic portal to New Physics, not just to dark matter.

It is a fact that SM neutrinos have mass, albeit tiny, as recently recognised by the 2015 Nobel prize for Physics. Many mechanisms and models have been proposed even before this discovery to explain the magnitude of neutrino masses. In the case of renormalisable models, neutrino masses arise when {\it some}\footnote{At present we do not measure directly neutrino masses but only two mass differences. Therefore, it is sufficient that only $2$ SM neutrinos are massive, and hence a minimum of $2$ RH neutrino should be introduced. Obviously, more than two can exist, if decoupled.} right-handed neutrinos $N$, singlets under the SM gauge groups, are introduced\footnote{Neutrino masses can be explained also when new scalar triplets are introduced, as in the so-called type-II seesaw model. Also there, mono-Higgs signatures can arise, of a different nature from those described here, i.e., without a fermionic mediator~\cite{Arhrib:2014nya}.}. Due to their nature of SM singlets, they are often called ``sterile'' neutrinos. In all possible models that comprise RH neutrinos, the latter will couple to the left-handed lepton doublet $L$ via renormalisable Yukawa interactions with the Higgs field $\phi$ such as
\begin{equation}\label{L_nuHN}
\mathcal{L} \ni \sum_i\, y^i_N\, \overline{L}\, \widetilde{\phi}^\dagger\, N^i\, .
\end{equation}
The above Lagrangian term gives neutrinos a Dirac mass and introduces couplings with the Higgs and would-be Goldstone bosons. However, to account for experimentally viable neutrino masses, the Yukawa couplings ought to be of $\mathcal{O}(10^{-12})$, unnaturally small. Being singlets of the SM, RH neutrinos can however also have a direct Majorana mass $M$, that can act as a counterpart to suppress the SM neutrino masses.
This is the case of the typical type-I/III seesaw models, where the smallness of neutrino masses is inversely related to the scale at which New Physics is introduced (the RH neutrino mass, $M$, typically at the GUT scale), yielding $m_\nu \sim y^2_N v^2_{EW}/M$. In essence, neutrino masses are suppressed by the ratio of coupling over energy scale, hence either by very large scales of New Physics or due to usually unmotivated small couplings.

In theories that comprise a protective ``lepton-number-like'' symmetry this is not the case anymore. A small breaking of this protective symmetry can give rise to the observed small neutrino masses. The latter are directly proportional to the amount of symmetry breaking $\epsilon$, as for instance in the case of the ``linear'' and ``inverse'' seesaw mechanisms. The main feature is the direct proportionality of the SM neutrino masses to the degree of symmetry breaking, $m_\nu \sim \epsilon\, y^2_N v^2_{EW}/M^2$, rather than to the scale at which New Physics is introduced. Ought to this, the magnitude of the $y^i_N$ couplings is not anymore connected to the scale of neutrino masses, that hence are free parameters that can be of $\mathcal{O}(1)$ even when New Physics is introduced at the TeV scale. In this case, when left-handed and right-handed neutrinos mix, yielding massive ``light'' and ``heavy'' neutrinos, the former are the SM-like ones, and the latter are at the TeV scale. Further,
%It is therefore possible to probe them at the LHC.
without specifying the nature of the underlying model, eq.~(\ref{L_nuHN}) yields the decay of heavy neutrinos into a SM-like neutrino and a neutral boson, or into a charged lepton and a $W$ boson, when kinematically allowed. In the large mass limit, the naive counting of degrees of freedom tells that the various branching ratios (BR) stay in a $BR(N\to\ell^\pm W^\mp) : BR(N\to\nu Z) : BR(N\to\nu H) \sim 2:1:1$ ratio. When allowed, the decay of a heavy neutrino into the Higgs boson and a SM neutrino yields a type of mono-Higgs signature that is not included in dark matter models: the {\it\underline{resonant}} mono-Higgs production. Despite in Ref.~\cite{Carpenter:2013xra} a similar case was briefly mentioned (i.e., when a $Z'$ boson mixed with the SM-Z boson decays into the latter, then yielding the missing energy when decaying into SM neutrinos, and a Higgs boson), the signature itself has not been studied in great details, and the case of a fermionic resonance has never been considered before. 

Inspired by models of neutrino masses with protective symmetries, where $M$ and $y^i_N$ are independent parameters, it is the aim of this paper to study in details the resonant mono-Higgs signature at the LHC in a simplified benchmark model, to fill the gaps in the mono-Higgs literature and to test the above benchmark model for its reinterpretation. We anticipate here that the resonant nature of the signature implies that the Higgs boson (and hence its decay products) is produced with larger transverse momenta than in typical dark matter cases. Furthermore, the associated missing transverse energy is also much larger, which is a key ingredient for its study. In general, one therefore expects that a higher degree of optimisation can be pursued to better extract the signal over the backgrounds. Finally, the resonant nature can also be exploited to gain access to the intermediate resonance yielding the signature under consideration.

This paper is organised as follows. In section~\ref{sect:model} we describe the benchmark model that will be used to derive the exclusion limits on the resonant mono-Higgs signature. Section~\ref{sect:LHCsensitivity} describe the LHC sensitivity to the mono-Higgs signature, analysed in two different Higgs decay modes, the $\gamma\gamma$ and the $b\overline{b}$ final states. Further, in section~\ref{sect:exclusions} we summarise the results and draw the exclusions in the neutrino benchmark model. Finally, we conclude in section~\ref{sect:conclusions}.

\section{Benchmark scenario for resonant mono-Higgs production}\label{sect:model}
As described in the Introduction, we supplement the literature of mono Higgs searches with a resonant channel. Inspired by models for neutrino mass generation, this resonance is a heavy neutrino.

In the following, we describe the benchmark model that is going to be used at the end for the reinterpretation of the results, the ``symmetry protected seesaw scenario'' (SPSS)~\cite{Antusch:2015mia}. The relevant feature encoded here is the possibility to have heavy neutrinos with masses around the electroweak (EW) scale together with unsuppressed (up to ${\cal O}$(1)) Yukawa couplings. Both are then independent parameters of the model.

In the SPSS, we consider a pair of sterile neutrinos $N_R^I$ $(I=1,2)$ and a suitable ``lepton-number-like'' symmetry where $N_R^1$ ($N_R^2)$ has the same (opposite) charge as the left-handed $SU(2)_L$ doublets $L^\alpha,\,\alpha=e,\mu,\tau$. The masses of the light neutrinos arise when this symmetry gets slightly broken.

The Lagrangian density in the symmetric limit is given by
\begin{equation}
\mathscr{L} = \mathscr{L}_\mathrm{SM} -  \overline{N_R^1} M N^{2\,c}_R - y_{\nu_{\alpha}}\overline{N_{R}^1} \widetilde \phi^\dagger \, L^\alpha+\mathrm{H.c.}\;,
\label{eq:lagrange}
\end{equation}
where $\mathscr{L}_\mathrm{SM}$ contains the usual SM field content and  $L^\alpha$ and $\phi$ being the lepton and Higgs doublets, respectively. The $y_{\nu_{\alpha}}$ are the complex-valued neutrino Yukawa couplings and the sterile neutrino mass parameter $M$ can be chosen as real without loss of generality.

It is assumed that the presence of further sterile neutrinos can be neglected. This is either because of their large masses, or tiny mixing couplings. In our particular case, one can assume that further sterile neutrinos have zero charge under the ``lepton-number-like'' symmetry, preventing them to couple or even mix with the other states. As benchmark scenario, we therefore consider only two sterile neutrinos as explained above.

For the study presented here, it is not restrictive to consider the neutrino mass matrix in the limit of exact symmetry. After EW symmetry breaking, the $5 \times 5$ mass matrix can be written as:
\begin{equation}
\mathscr{L}_{\rm mass} = -\frac{1}{2} \left(\begin{array}{c} \overline{\nu^c_{e_L}} \\ \overline{\nu^c_{\mu_L}} \\ \overline{\nu^c_{\tau_L}} \\ \overline{N_R^1} \\ \overline{N_R^2} \end{array}\right)^T\,
\left( \begin{array}{ccccc}  0 & 0 & 0 & m_e & 0 \\ 0 & 0 & 0 & m_\mu & 0 \\ 0 & 0 & 0 & m_\tau & 0 \\ m_e &  m_\mu & m_\tau & 0 & M \\ 0 & 0 & 0 & M & 0 \end{array}\right) \left(\begin{array}{c} \nu_{e_L}\\\nu_{\mu_L}\\\nu_{\tau_L}\\ \left(N_R^1\right)^c\\ \left(N_R^2\right)^c \end{array} \right) +\mathrm{H.c.}\,,
\label{eq:massmatrix}
\end{equation}
with the Dirac masses $m_\alpha = y_{\nu_\alpha} v_\mathrm{EW}/\sqrt{2}$, where $v_\mathrm{EW}=246$ GeV is the SM vacuum expectation value. For sake of completeness, 2 small neutrino masses of ${\cal O}(\epsilon \,|y_{\nu_i}|^2 v^2_\mathrm{EW}/M^2)$ arise when the protective symmetry gets suitably slightly broken by $\epsilon \gtrsim 0$, while the third SM-like neutrino stays massless. However, a further sterile state could be added to give mass to this neutrino as well.

%The parameter $\epsilon$ controls the explicit breaking of the ``lepton-number-like'' symmetry. In the limit of exact symmetry, i.e.\ $\epsilon \equiv 0$, the three lightest neutrinos are massless, and the two heavy neutrinos are mass degenerate. Conversely, non-zero masses for the light neutrinos of ${\cal O}(\epsilon \,|y_{\nu_i}|^2 v^2_\mathrm{EW}/M^2)$ arise when the protective symmetry gets slightly broken by $\epsilon \gtrsim 0$.

The diagonalisation of the mass matrix from eq.~(\ref{eq:massmatrix}) results in the five mass eigenstates of the three light (possibly massless) and two heavy neutrinos and defines the unitary $5\times5$ leptonic mixing matrix $U$. 
With the active sterile mixing angles defined as
\begin{equation}
\theta_\alpha = \frac{y_{\nu_\alpha}^{*}}{\sqrt{2}}\frac{v_\mathrm{EW}}{M}\,,
\label{def:thetaa}
\end{equation}
the leptonic mixing matrix $U$ (unitary up to second order in $\theta_\alpha$ and neglecting the tiny perturbation from the light neutrino masses) is given by
\begin{equation} 
U = \left(\begin{array}{ccccc} 
{\cal N}_{e1}	& {\cal N}_{e2}	& {\cal N}_{e3}	& - \frac{\mathrm{i}}{\sqrt{2}}\, \theta_e & \frac{1}{\sqrt{2}} \theta_e 	\\ 
{\cal N}_{\mu 1}	& {\cal N}_{\mu 2}  	& {\cal N}_{\mu 3}  	& - \frac{\mathrm{i}}{\sqrt{2}}\theta_\mu & \frac{1}{\sqrt{2}} \theta_\mu  \\
{\cal N}_{\tau 1}	& {\cal N}_{\tau 2} 	& {\cal N}_{\tau 3} 	& - \frac{\mathrm{i}}{\sqrt{2}} \theta_\tau & \frac{1}{\sqrt{2}} \theta_\tau \\  
0	   	& 0		& 0	&  \frac{ \mathrm{i}}{\sqrt{2}} & \frac{1}{\sqrt{2}}\\
-\theta^{*}_e	   	& -\theta^{*}_\mu	& -\theta^{*}_\tau &\frac{-\mathrm{i}}{\sqrt{2}}(1-\tfrac{1}{2}\theta^2) & \frac{1}{\sqrt{2}}(1-\tfrac{1}{2}\theta^2)
\end{array}\right)\,.
\label{eq:mixingmatrix}
\end{equation}
The elements of the $3\times 3$ submatrix ${\cal N}$, which is the effective mixing matrix of the three active neutrinos, i.e. the Pontecorvo--Maki--Nakagawa--Sakata (PMNS) matrix, are given as
\begin{equation}\label{eq:matrixN}
{\cal N}_{\alpha i} = (\delta_{\alpha \beta} - \tfrac{1}{2} \theta_{\alpha}\theta_{\beta}^*)\,(U_\ell)_{\beta i}\,,
\end{equation}
with $U_\ell$ being a unitary $3 \times 3$ matrix. It is clear that the submatrix ${\cal N}$ is non-unitary.\\

This benchmark model can contribute to the resonant Higgs + MET signature by singly-producing a heavy neutrino, that will decay into a Higgs boson and a SM-like neutrino. A single heavy neutrino is produced via an off-shell $Z^\ast$ or $W^{\pm\ast}$, in association with a SM-like neutrino or a charged lepton, respectively, as in figure~\ref{fig:LHChiggsproduction}. The diagram on the right hand, the production in association with a charged lepton, can give a Higgs + MET signature when the charged lepton is not reconstructed. For the reinterpretation in section~\ref{sect:exclusions}, we set
\begin{equation}
y_{\nu_\alpha} \equiv y_{\nu} \qquad \forall \alpha=e,\,\mu,\,\tau.
\end{equation}

\FIGURE{
\centering
\includegraphics[width=0.45\textwidth]{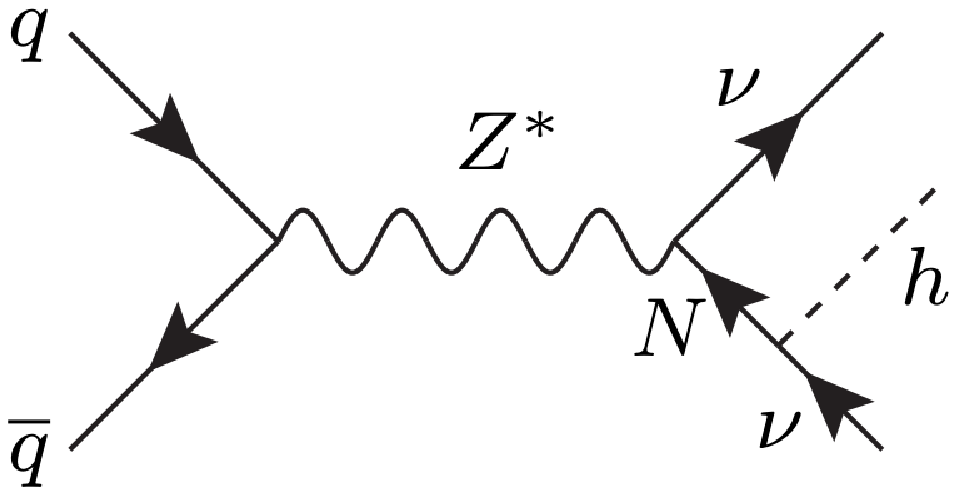}\hspace{0.5cm}
\includegraphics[width=0.45\textwidth]{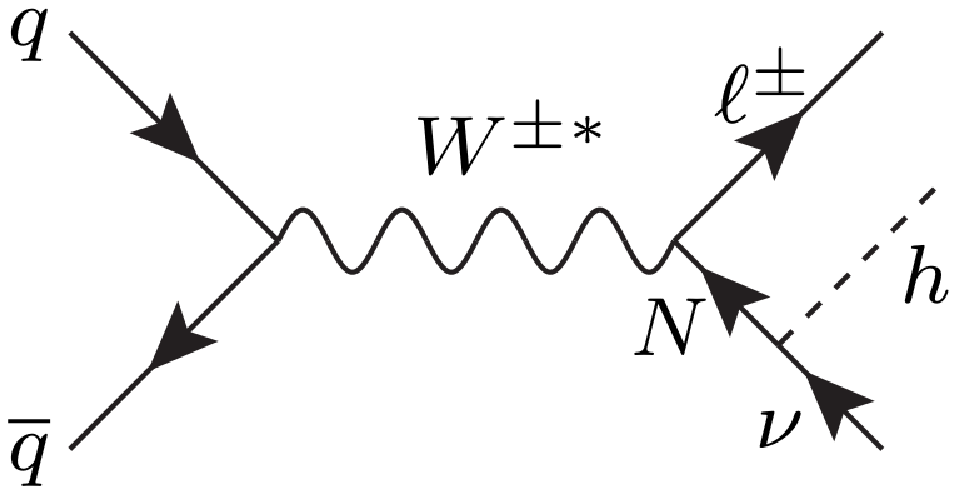}
\caption{LHC production mechanisms for the resonant $H+$MET signature.\label{fig:LHChiggsproduction}}
}

Parton level cross sections for these processes in the benchmark model are as in figure~\ref{fig:xs}. They will be employed in section~\ref{sect:reinterpretation} to reinterpret the exclusion bounds.

\FIGURE{
\centering
\includegraphics[width=0.75\textwidth]{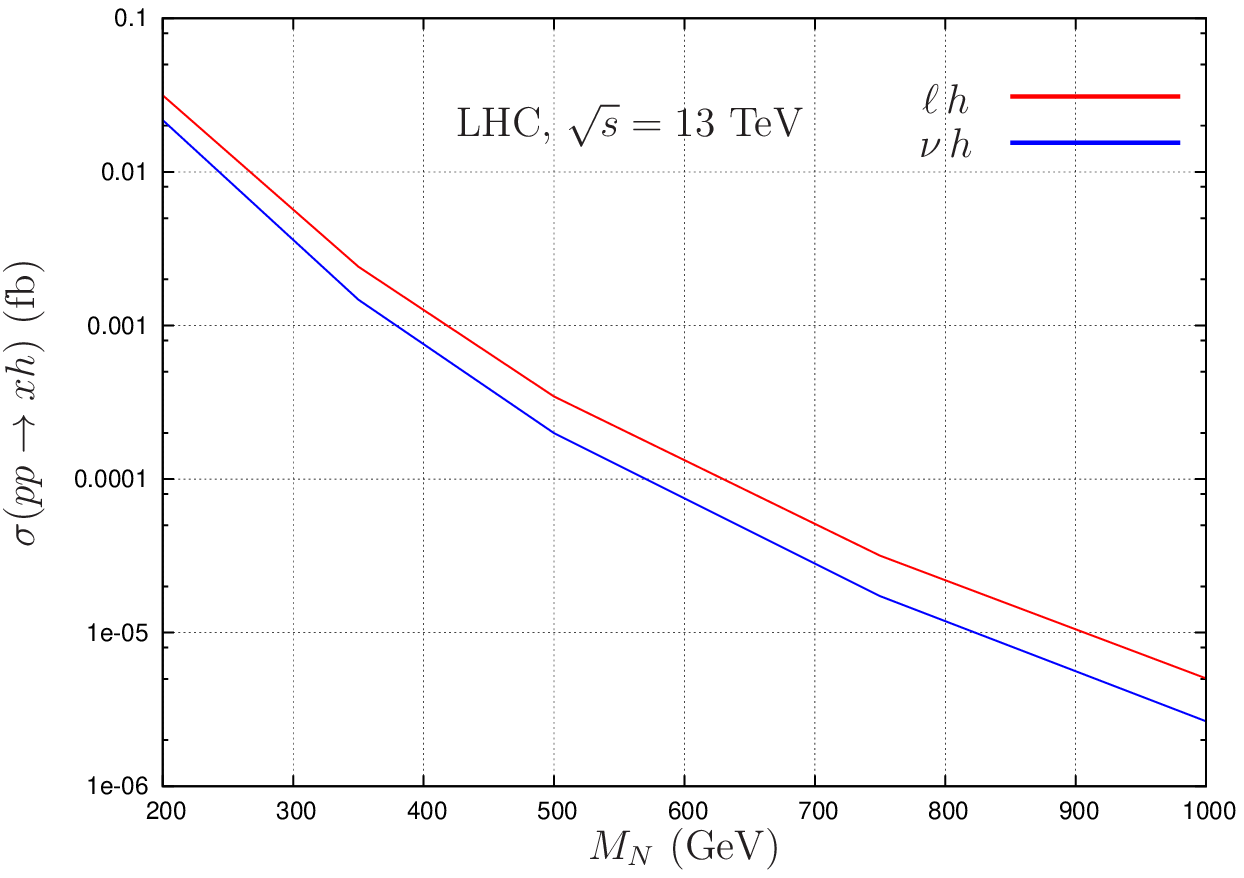}
\caption{Benchmark model cross sections for the processes of figure~\ref{fig:LHChiggsproduction}, for $y_{\nu}=0.01$.\label{fig:xs}}
}

\subsection{Summary of existing constraints}
Before moving to the analysis of the LHC sensitivities to the Higgs +MET signature, we briefly review the present exclusion bounds on the parameters of the SPS model. The presence of heavy neutrinos could manifest both in direct and indirect measurements. For a recent review, see Refs.~\cite{Antusch:2014woa,Antusch:2015mia}.

If sufficiently light (i.e., below $\sim 200$ GeV),sterile neutrinos could be directly produced at colliders. The LEP experiments have searched for them in a variety of channels. Those yielding the tightest bounds are the searches for neutrinos produced at the $Z$ pole by DELPHI~\cite{Abreu:1996pa}, and for modification to the $e^+ e^− \to W^+ W^−$ SM process, both in the $\ell^+ \ell^{'−} +$ MET and $\ell + 2j+$ MET channels, at ALEPH~\cite{Heister:2004wr} and L3~\cite{Achard:2001qv}, respectively.

Even when too heavy to be directly produced, sterile states indirectly affect other measurements. The non-unitarity of the PMNS (sub)matrix, see eq.~\ref{eq:matrixN}, will change the weak currents, as well as modifying the relation between the Fermi constant and the muon decay rate. The most sensitive indirect constraints typically come from the electroweak precision observables (EWPO), lepton universality, rare flavour-violating charged lepton
decays ($\ell^\alpha \to \ell^\beta \gamma$), and unitarity of CKM.

From a global analysis of all observables that can be sensitive to the presence of sterile neutrinos, the following conservative upper bounds at $68\%$ ($1\sigma$) Bayesian confidence level are obtained~\cite{Antusch:2014woa,Antusch:2015mia}:
\begin{eqnarray}\label{eq:bounde}
y_{\nu_e} &\leq& 0.042 \cdot \frac{M_N/GeV}{175}, \\ \label{eq:boundmu}
y_{\nu_\mu} &\leq& 0.065 \cdot \frac{M_N/GeV}{175}, \\ \label{eq:boundtau}
y_{\nu_\tau} &\leq& 0.015 \cdot \frac{M_N/GeV}{175}, \\
\end{eqnarray}

It is worth noticing, that the presence of active-sterile mixing is not only bounded by the precision data, but can even improve on the global fit, in particular when a selective set of observables is chosen~\cite{Basso:2013jka}.

\section{LHC sensitivity}\label{sect:LHCsensitivity}
In this section we discuss the sensitivity of the resonant mono-Higgs signature at the LHC run-II, for $\sqrt{s}=13$ TeV and $\mathcal{L}=100$ fb$^{-1}$ integrated luminosity. Prospects for the High Luminosity phase, where $3$ ab$^{-1}$ of data are expected to be collected, will also be given. Before outlining the analyses strategies, we describe the simulation environment.

All samples employed in this study have been generated in {\tt MadGraph5\_aMC@NLO~v2.1.2}~\cite{Alwall:2014hca} with the {\tt CTEQ6L1} PDF~\cite{Pumplin:2002vw}. Events have subsequently been hadronised/parton showered in {\tt PYTHIA~6}~\cite{Sjostrand:2006za} with tune Z2~\cite{Field:2011iq}. Detector simulation is performed with a customised version of {\tt Delphes~3}~\cite{deFavereau:2013fsa} to emulate the CMS detector. Jets have been reconstructed with {\tt FastJet}~\cite{Cacciari:2011ma} employing the anti-$k_t$ algorithm~\cite{Cacciari:2008gp} with parameter $R=0.5$.

The signal (S) is generated at leading order from the model implemented in {\tt FeynRules}~\cite{Alloul:2013bka}.
We generated $5$ benchmark points for $N_1$ neutrino mass as 
\begin{equation}
M \in \left[ 200,\,350,\,500,\,750,\,1000\right] \mbox{ GeV,}
\end{equation}
at leading order plus up to one merged jet. The cross sections for each sample is left as arbitrary for setting the LHC sensitivities. For the reinterpretation of the sensitivities in the SPSS, the samples will be normalised to the partonic cross sections of figure~\ref{fig:xs} without applying any $k$-factor, adopting a more conservative approach.

The analyses that will follow are distinguished by the final state decay products of the Higgs boson, either $h\to b\overline{b}$ or $h\to \gamma\gamma$. The various backgrounds of each analysis will be outlined later. In general, we generated leading order samples with up to 2 merged jets normalised to the (N)NLO cross section where available, taken from~\cite{Alwall:2014hca,Czakon:2013goa}.
For both signal and background, a suitable number of unweighted event is generated. 
%We checked that the statistical uncertainties at any point of our analyses are below $1\%$, and for this reason they will not be quoted.
We did not simulate multijet backgrounds, which can be reliably estimated only from data. It will be shown anyway that such backgrounds can be safely ignored. 

The analysis is carried out in {\tt MadAnalysis~5}~\cite{Conte:2012fm,Conte:2014zja}.
Photons and jets are identified if passing the following criteria:
\begin{eqnarray}
p_T(\gamma) > 30 {\mbox{ GeV,}} &\qquad & |\eta(\gamma)| < 2.5\, ,\\
p_T(j) > 40 {\mbox{ GeV,}} &\qquad & |\eta(j)| < 3\, , \\
\Delta R(\gamma, j) > 0.4, &\qquad & 
\end{eqnarray}
while isolated ``loose'' charged leptons, i.e., with $p^\ell_T > 10$ GeV, $|\eta^{\ell}| < 2.5(2.4)$ for $\ell = e(\mu)$, are selected for vetoing.

External routines for $b$-tagging and for lepton and photon isolation have been implemented. Regarding the former, here we adopted the medium working point~\cite{Chatrchyan:2012jua}, which has an average $b$-tagging rate of $70\%$ and a light mistag rate of $1\%$. To apply the $b$-tagging, we considered jets within the tracker only, i.e. with $|\eta(j)| < 2.4$. This means that the $b$-tagging probability for jets with larger pseudorapidities is vanishing. For the latter, the combined tracker-calorimetric isolation is used to identify isolated leptons/photons. The relative isolation $I_{rel}$ is defined as the sum of the $p_T$ and calorimetric deposits of all tracks within a cone of radius $\Delta R=0.4$, divided by the $p_T$ of the lepton/photon. The latter is isolated if $I_{rel}\leq 0.20$ for loose leptons and $I_{rel}\leq 0.10$ for photons.

After the object reconstruction and selection, we apply some general preselections as follows: we require at least 2 b-tagged jet or photons, and no other jet nor loose leptons (electrons or muons).

%Efficiencies and event yields are evaluated for $\mathcal{L}=100$ fb$^{-1}$ and are collected in table~\ref{tab:presel_eff}. 

We describe in the following the analyses differentiating the two selected decay channel for the Higgs boson: $h\to b\overline{b}$ and $h\to \gamma\gamma$, that are the two most sensitive channels~\cite{Carpenter:2013xra}.

%%%%%%%%%%%%5%%%%%%%%%%%%5%%%%%%%%%%%%5%%%%%%%%%%%%5%%%%%%%%%%%%5
%%%%%%%%%%%%5%%%%%%%%%%%%5%%%%%%%%%%%%5%%%%%%%%%%%%5%%%%%%%%%%%%5
%			h -> bb
%%%%%%%%%%%%5%%%%%%%%%%%%5%%%%%%%%%%%%5%%%%%%%%%%%%5%%%%%%%%%%%%5
%%%%%%%%%%%%5%%%%%%%%%%%%5%%%%%%%%%%%%5%%%%%%%%%%%%5%%%%%%%%%%%%5

\subsection{$h\to b\overline{b}$}
The first final state we consider is particularly interesting since typically very challenging. At the LHC, this decay mode for the SM Higgs can be accessed in all channels but in gluon fusion. However, the presence of large MET in the final state here considered proves to be essential to access also the latter. These events can be triggered for instance with a MET trigger, that requires MET $\geq 200$ GeV to be fully functional. We remind that as object selection, we require the presence of exactly 2 jets, both b-tagged, and that we veto the presence of any charged loose lepton.
 
Backgrounds to the $b\overline{b}+$ MET signature are
\begin{itemize}
\item $gg\to H\to b\overline{b}$, the SM Higgs production via gluon fusion, labelled $H$;
\item $W^\pm H/ZH$ with $H\to b\overline{b}$, the SM Higgs strahlung production, labelled $WH/ZH$;
\item $Zb\overline{b}/W^\pm b\overline{b}$, Drell-Yan processes plus 2 extra $b$-jets, labelled $W/Z+jets$;
\item single-top processes, labelled $T+jets$;
\item $t\overline{t}$, top pair production, labelled $TT+jets$.
\end{itemize}

As mentioned earlier on, the presence of a large missing energy in the mono-Higgs signature is essential to capture this signal, since a MET-based trigger can be employed, requiring
\begin{equation}\label{eq:MET-cut}
{\mbox{MET}} \geq 200 {\mbox{ GeV}}
\end{equation}
to be in the plateau. This request completely suppresses the QCD multijet background, that we did not simulate, and the SM Higgs boson production via gluon fusion, see table~\ref{Table:CutFlow-sgn}. In figure~\ref{fig:MET-Mbb}(left) we see that the signal suffers this selection only for low resonance masses, while most of the backgrounds are already heavily suppressed. Notice that the signal is plotted for a nominal cross section of $1$~pb for a better comparison. In figure~\ref{fig:MET-Mbb}(right) we plot the reconstructed Higgs mass. Backgrounds that are non-resonant in the $b\overline{b}$ system have larger tails, in particular $t\overline{t}$. To suppress them, a cut in the $b\overline{b}$ invariant mass is selected as follows:
\begin{equation}\label{eq:MH-cut}
100 \leq M(b\overline{b})/\mbox{GeV} \leq 150\, ,
\end{equation}
best optimised to enhance the signal-over-background ratio compatibly with the Higgs mass resolution in the dijet final state.
\FIGURE{
\centering
\includegraphics[width=0.48\textwidth]{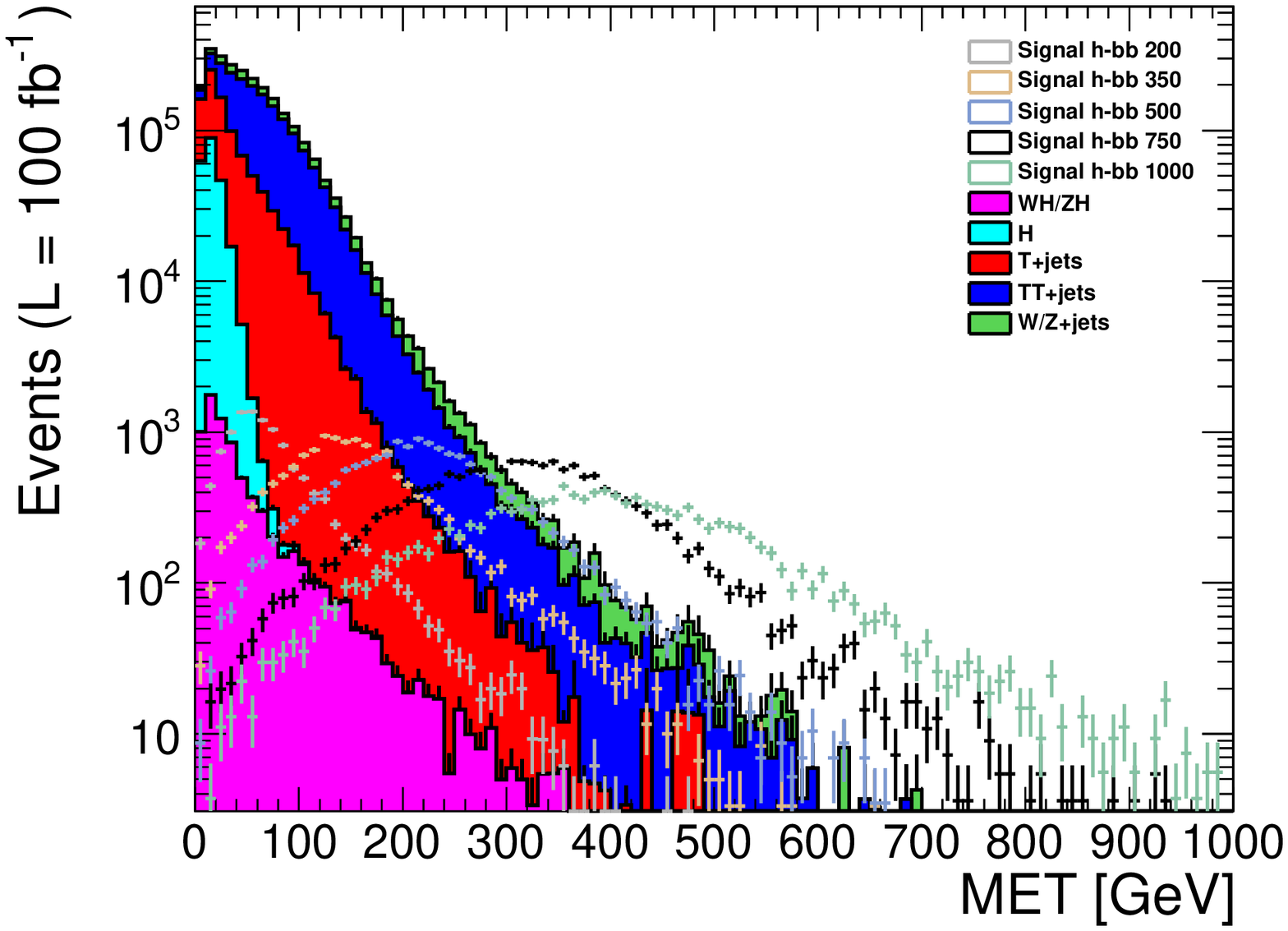}
\includegraphics[width=0.48\textwidth]{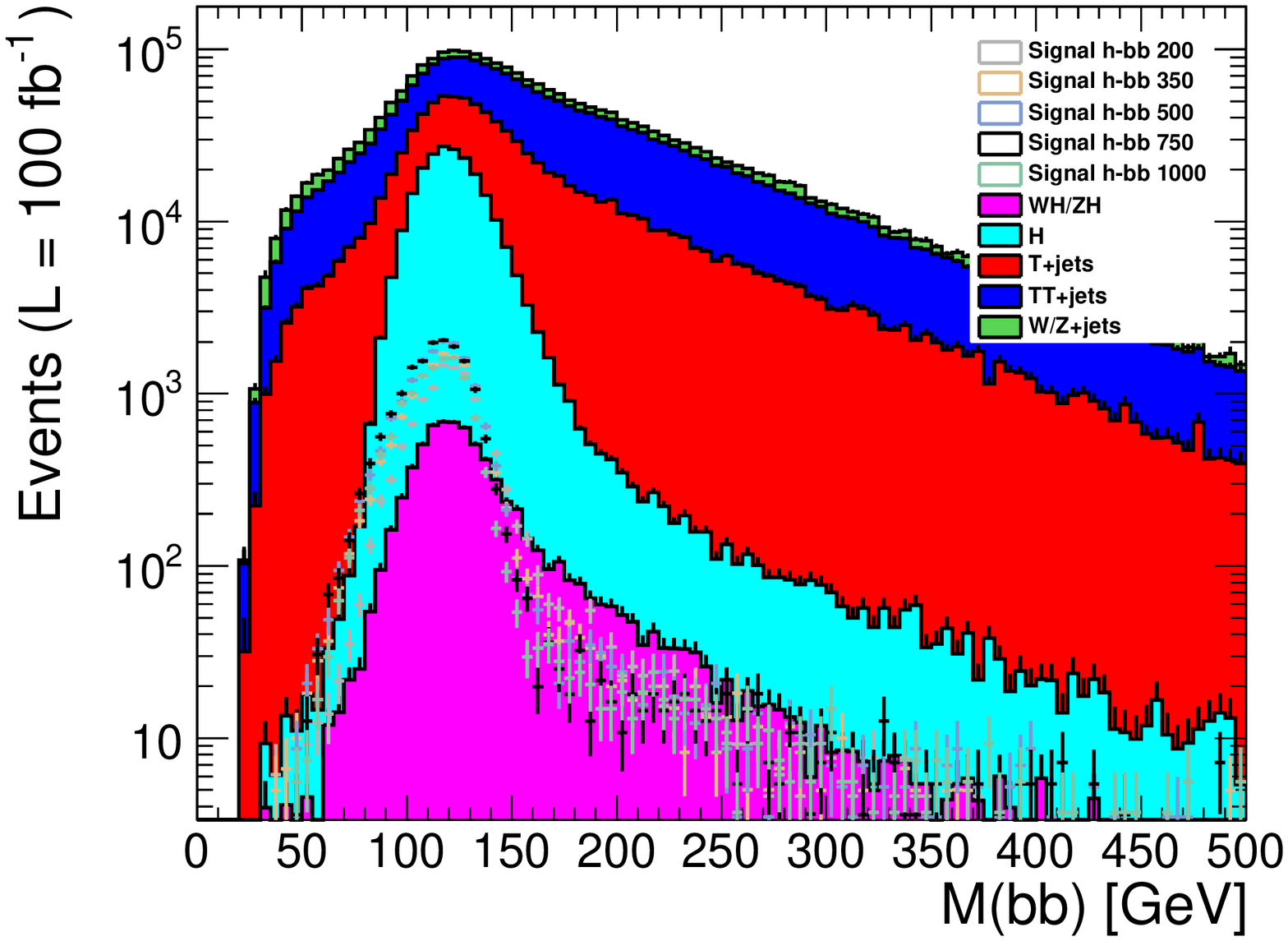}
\caption{(Left) MET, missing transverse energy. (Right) $b\overline{b}$ invariant mass. Signal plotted for a nominal cross section of $1$~pb.\label{fig:MET-Mbb}}
}

The initial selection of MET $\geq 200$ GeV, dictated by the trigger, implies that the Higgs boson in the signal has large $p_T$. This can be seen in figure~\ref{fig:PTh-MTbb}(left), drawn after selecting the Higgs mass window as in eq.~(\ref{eq:MH-cut}). It can be seen that the Higgs boson in the signal has mostly to have a transverse momentum larger than $200$ GeV as well, for momentum conservation. Hence, we can reinforce this statement applying the following cut:
\begin{equation}\label{eq:ptH-cut}
p_T(b\overline{b}) \geq 200 \mbox{ GeV}\, ,
\end{equation}
that will affect almost only the background. 

\FIGURE{
\centering
\includegraphics[width=0.48\textwidth]{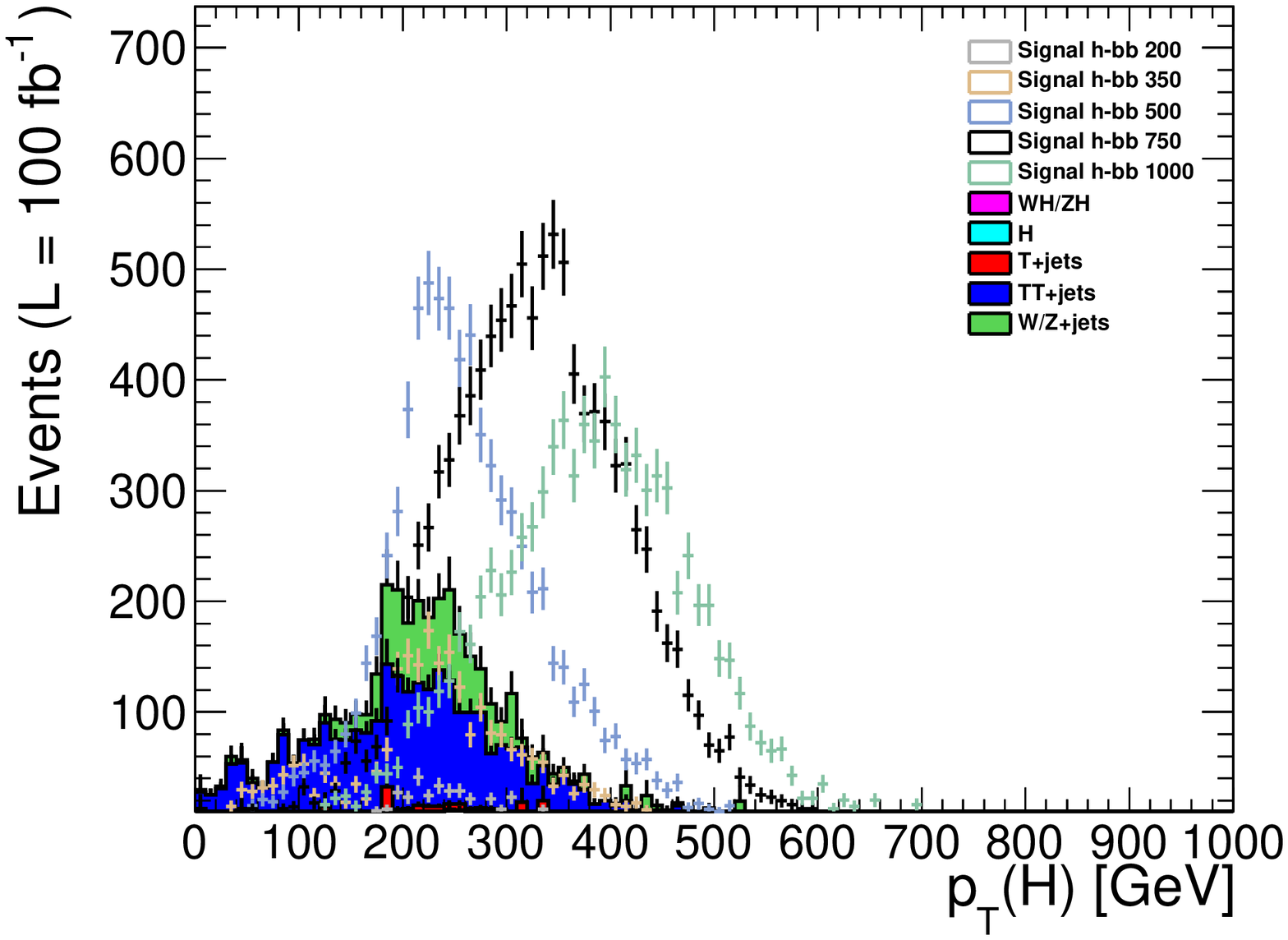}
\includegraphics[width=0.48\textwidth]{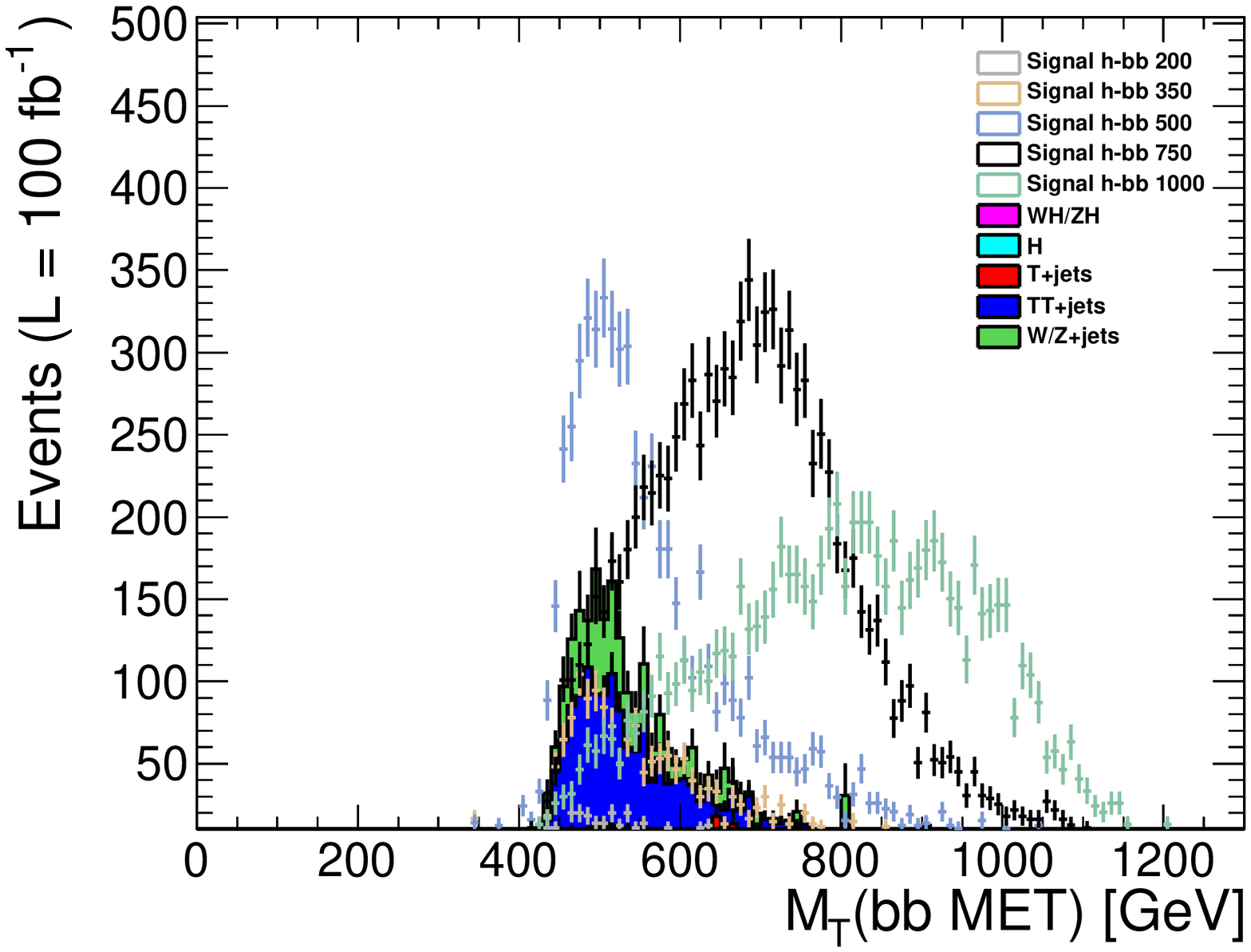}
\caption{(Left) Transverse momentum of the Higgs boson, after cut~(\ref{eq:MH-cut}). (Right) Transverse mass of total  Higgs $+$ MET system. Signal plotted for a nominal cross section of $1$~pb.\label{fig:PTh-MTbb}}
}

Lastly, figure~\ref{fig:PTh-MTbb}(right) shows that despite the final state has 2 sources of missing energy (2 light SM-like neutrinos), the transverse mass of the whole final state ($b\overline{b}+$ MET) is sensitive to the intermediate resonance mass. This is due to the charged-current production mechanism as in figure~\ref{fig:LHChiggsproduction}(right), where the intermediate resonance yields the only source of missing energy, on top of a broader behaviour from the neutral current production.
However, the application of eqs.~(\ref{eq:MET-cut})--(\ref{eq:ptH-cut}) will shape this variable, that is suppressed for values lower than $\simeq 400$ GeV. It however stays meaningful for benchmarks with $M_N \geq 500$ GeV. In these cases (for the last 2 benchmark points we simulated), a further selection can be applied, that is optimised to best enhance the signal-over-background ratio.

Efficiencies of the various cuts for the backgrounds and for the signal benchmarks (for a nominal normalisation cross section of $1$ pb) are collected in tables~\ref{Table:CutFlow-bkg}--\ref{Table:CutFlow-sgn}. 

\begin{table}[!h]
\scalebox{0.6}{
\begin{tabular}{|l|c||c|c|c|c|c|c|c|c|}
\hline
\hline
Cut & Sum Backgrounds  &  $WH/ZH$  & $H$  & $T+jets$  & $TT+jets$  & $W/Z+jets$ \\
\hline
no cuts  & $(18.656 \pm 0.003)\cdot 10^{9}$ & 126199 $\pm$ 317 & 1857012 $\pm$ 1590 & 28105716 $\pm$ 16855 & 39761760 $\pm$ 7677 & $(18.586 \pm 0.003) \cdot 10^{9}$ \\
$N_{\ell} \equiv 0 $ &  $(18.607 \pm 0.003)\cdot 10^{9}$ & 122114 $\pm$ 311 & 1856968 $\pm$ 1589 & 26173762 $\pm$ 16392 & 33606184 $\pm$ 7147 &  $(18.545 \pm 0.003)\cdot 10^{9}$ \\
$N_j \equiv 2 $ & ($146.59 \pm 0.09)\cdot 10^6$ & 38034 $\pm$ 174 & 684088 $\pm$ 957 & 8604696 $\pm$ 9307 & 8947317 $\pm$ 3779 & ($128.32 \pm 0.09)\cdot 10^6$ \\
$N_b \equiv 2 $ & 2513227 $\pm$ 5215 & 7353 $\pm$ 75 & 216998 $\pm$ 540 & 719272 $\pm$ 2808 & 1317921 $\pm$ 1366 & 251682 $\pm$ 4141 \\
MET $\geq 200  $ & 21914 $\pm$ 325 & 173 $\pm$ 11 & 0 $\pm$ 0 & 2064 $\pm$ 141 & 12921 $\pm$ 134 & 6755 $\pm$ 260 \\
$100<M(b\overline{b})/\mbox{GeV}<150  $ & 4042 $\pm$ 105 & 141 $\pm$ 10 & 0 $\pm$ 0 & 201 $\pm$ 35 & 2637 $\pm$ 64 & 1063 $\pm$ 74 \\
$p_T(H) \geq 200$ GeV & 2386 $\pm$ 83 & 130 $\pm$ 10 & 0 $\pm$ 0 & 111 $\pm$ 19 & 1373 $\pm$ 49 & 772 $\pm$ 64 \\
\hline
\hline
\end{tabular}
}
\caption{ CutFlow for backgrounds for $b\overline{b}+$MET, events for $\mathcal{L}=100$ fb$^{-1}$.}
\label{Table:CutFlow-bkg}
\end{table}

\begin{table}[!h]
\scalebox{0.75}{
\begin{tabular}{|l|c|c|c|c|c|c|}
\hline
\hline
Cut & $h\ell$@200 GeV & $h\nu$@200 GeV & $h\ell$@350 GeV & $h\nu$@350 GeV &  $h\ell$@500 GeV & $h\nu$@500 GeV \\
\hline
no cuts  & 99969 $\pm$ 392 & 99969 $\pm$ 391 &99980 $\pm$ 408 & 99979 $\pm$ 407 & 99986 $\pm$ 417 & 99986 $\pm$ 416\\
$N_{\ell} \equiv 0 $  & 89281 $\pm$ 370 & 96682 $\pm$ 385 & 93397 $\pm$ 394 & 97390 $\pm$ 402 &  95405 $\pm$ 407 & 97995 $\pm$ 412  \\
$N_j \equiv 2 $ & 25389 $\pm$ 197 & 38034 $\pm$ 174 & 32708 $\pm$ 233 & 33394 $\pm$ 235 & 35630 $\pm$ 249 & 37583 $\pm$ 255  \\
$N_b \equiv 2 $ & 5354 $\pm$ 91 & 6239 $\pm$ 98 & 7611 $\pm$ 112 & 7353 $\pm$ 75 & 7406 $\pm$ 114 & 8952 $\pm$ 125 \\
MET $\geq 200  $  & 61 $\pm$ 10 & 453 $\pm$ 26 & 964 $\pm$ 40 & 2264 $\pm$ 61 & 3968 $\pm$ 83 & 5643 $\pm$ 99 \\
$100<M(b\overline{b})/\mbox{GeV}<150  $ & 43 $\pm$ 8 & 357 $\pm$ 23 & 737 $\pm$ 35 & 1778 $\pm$ 54 & 3011 $\pm$ 72 & 4557 $\pm$ 89 \\
$p_T(H) \geq 200$ GeV & 31 $\pm$ 7 & 302 $\pm$ 21 & 258 $\pm$ 21 & 1524 $\pm$ 50 & 1980 $\pm$ 59 & 4207 $\pm$ 85 \\
\hline
\end{tabular}
}
\scalebox{0.75}{
\begin{tabular}{|l|c|c|c|c|c|}
\hline
Cut & Sum Backgrounds  & $h\ell$@750 GeV & $h\nu$@750 GeV & $h\ell$@1 TeV  & $h\nu$@1 TeV\\
\hline
no cuts  & $(18.656 \pm 0.003)\cdot 10^{9}$  & 99992 $\pm$ 425 & 99992 $\pm$ 424 & 99996 $\pm$ 431 & 99996 $\pm$ 431 \\
$N_{\ell} \equiv 0 $ &  $(18.607 \pm 0.003)\cdot 10^{9}$ & 96793 $\pm$ 418 & 98690 $\pm$ 422 & 97487 $\pm$ 425 & 99044 $\pm$ 429 \\
$N_j \equiv 2 $ & ($146.59 \pm 0.09)\cdot 10^6$ & 37225 $\pm$ 259 & 39248 $\pm$ 266 & 36516 $\pm$ 260 & 37387 $\pm$ 263  \\
$N_b \equiv 2 $ & 2513227 $\pm$ 5215 & 7390 $\pm$ 115 & 9198 $\pm$ 129 & 5838 $\pm$ 104 & 7025 $\pm$ 114 \\
MET $\geq 200  $ & 21914 $\pm$ 325  & 6255 $\pm$ 106 & 7972 $\pm$ 120 & 5337 $\pm$ 100 & 6424 $\pm$ 109 \\
$100<M(b\overline{b})/\mbox{GeV}<150  $ & 4042 $\pm$ 105 & 4710 $\pm$ 92 & 6171 $\pm$ 105 & 3840 $\pm$ 84 & 4883 $\pm$ 95  \\
$p_T(H) \geq 200$ GeV & 2386 $\pm$ 83  & 4217 $\pm$ 87 & 6016 $\pm$ 104 & 3626 $\pm$ 82 & 4814 $\pm$ 95  \\
\hline
$ M_T (b\overline{b}$ MET$) \geq 640$ GeV & 479 $\pm$ 41 & 2830 $\pm$ 71 & 3845 $\pm$ 83 & $-$ & $-$  \\
$ M_T (b\overline{b}$ MET$) \geq 820$ GeV & 80 $\pm$ 12  & $-$ & $-$ & 2045 $\pm$ 62 & 2119 $\pm$ 63  \\
\hline
\hline
\end{tabular}
}
\caption{ CutFlow for signals for $b\overline{b}+$MET, events for $\mathcal{L}=100$ fb$^{-1}$ for a nominal cross section of $1$ pb.}
\label{Table:CutFlow-sgn}
\end{table}

We define the significance $\mathbb{S}$ of each signal sample as follows:
\begin{equation}\label{eq:significance}
\mathbb{S} = \frac{S}{\sqrt{S+B}}\,,
\end{equation}
where $S$ and $B$ are the number of events for signal and background, respectively.
From the surviving signal events, one can deduced the value of the initial cross section for the various signal benchmark points to achieve a significance of $2\sigma$, commonly understood as ``excluded'' if no signal is observed. The excluded cross sections evaluated as discussed are collected in table~\ref{tab:xsExcl-bb} and displayed in Figure~\ref{fig:xsExcl}.

\begin{table}[!h]
\centering
\scalebox{0.8}{
\begin{tabular}{|l|c|c|c|c|c|}
\hline
\hline
$h\to b\overline{b}$ & $M=200$ GeV  & $M=350$ GeV  & $M=500$ GeV  & $M=750$ GeV  & $M=1000$ GeV \\
\hline
$\sigma$ (fb), $\mathcal{L}=100$ fb$^{-1}$ & $704.7\pm 44.4$ & $135.2\pm 3.9$ & $35.69\pm 0.57$ & $14.37\pm 0.21$ &$9.65\pm 0.17$\\
$\sigma$ (fb), $\mathcal{L}=3$ ab$^{-1}$ & $126.5\pm 8.0$ & $24.3\pm 0.7$ & $6.41\pm 0.10$ & $2.53\pm 0.04$ &$1.61\pm 0.03$\\
\hline
\hline
\end{tabular}
}
\caption{ Excluded cross sections in the $b\overline{b}$ final state, at the LHC for $\sqrt{s}=13$ TeV.}
\label{tab:xsExcl-bb} 
\end{table}

A final remark is in place. Large neutrino masses means that the Higgs has a boost. From a naive expectation, $\Delta R \simeq 2m / p_T$, when the Higgs boson is produced with $p_T$ larger than $500$ GeV, the 2 $b$-jets in the final state will be typically closer than $\Delta R = 0.5$, that is our jet cone. Meaning, the 2 jets will start to merge and look like a single one. Hence, for heavy neutrino masses larger than 1 TeV, for which in most of the cases the Higgs boson has a $p_T$ larger than $500$ GeV, a simple dijet final state analysis might not be efficient enough to single out our signal. In this case one might investigate how more involved boosted techniques can improve the signal-over-background discrimination.

%%%%%%%%%%%%5%%%%%%%%%%%%5%%%%%%%%%%%%5%%%%%%%%%%%%5%%%%%%%%%%%%5
%%%%%%%%%%%%5%%%%%%%%%%%%5%%%%%%%%%%%%5%%%%%%%%%%%%5%%%%%%%%%%%%5
%			h -> aa
%%%%%%%%%%%%5%%%%%%%%%%%%5%%%%%%%%%%%%5%%%%%%%%%%%%5%%%%%%%%%%%%5
%%%%%%%%%%%%5%%%%%%%%%%%%5%%%%%%%%%%%%5%%%%%%%%%%%%5%%%%%%%%%%%%5

\subsection{$h\to \gamma\gamma$}
The diphoton final state is the most sensitive one for the SM Higgs boson at the LHC and for the Higgs + MET signature in dark matter models, see Ref.~\cite{Carpenter:2013xra}. This very clean signal does not require any MET selection for triggering. However, a lower value can be selected for enhancing the signal.
We remind that as object selection, we require the presence of exactly 2 photons, and that we veto the presence of any jet or charged loose lepton.
 
Backgrounds to the $\gamma\gamma+$ MET signature are
\begin{itemize}
\item $gg\to H\to \gamma\gamma$, the SM Higgs production via gluon fusion, labelled $H$;
\item $W^\pm H/ZH$ with $H\to \gamma\gamma$, the SM Higgs strahlung production, labelled $WH/ZH$;
\item $pp\to\gamma\gamma$, the SM diphoton production, labelled $AA$;
\item $pp\to\gamma\gamma Z/W^\pm$, the SM diphoton production plus a vector boson that supplies some further MET, labelled $AAV$.
\end{itemize}

In figure~\ref{fig:MET-Maa} the MET and the diphoton invariant mass distributions can be seen on the left panel and on the right panel respectively.

\FIGURE{
\centering
\includegraphics[width=0.48\textwidth]{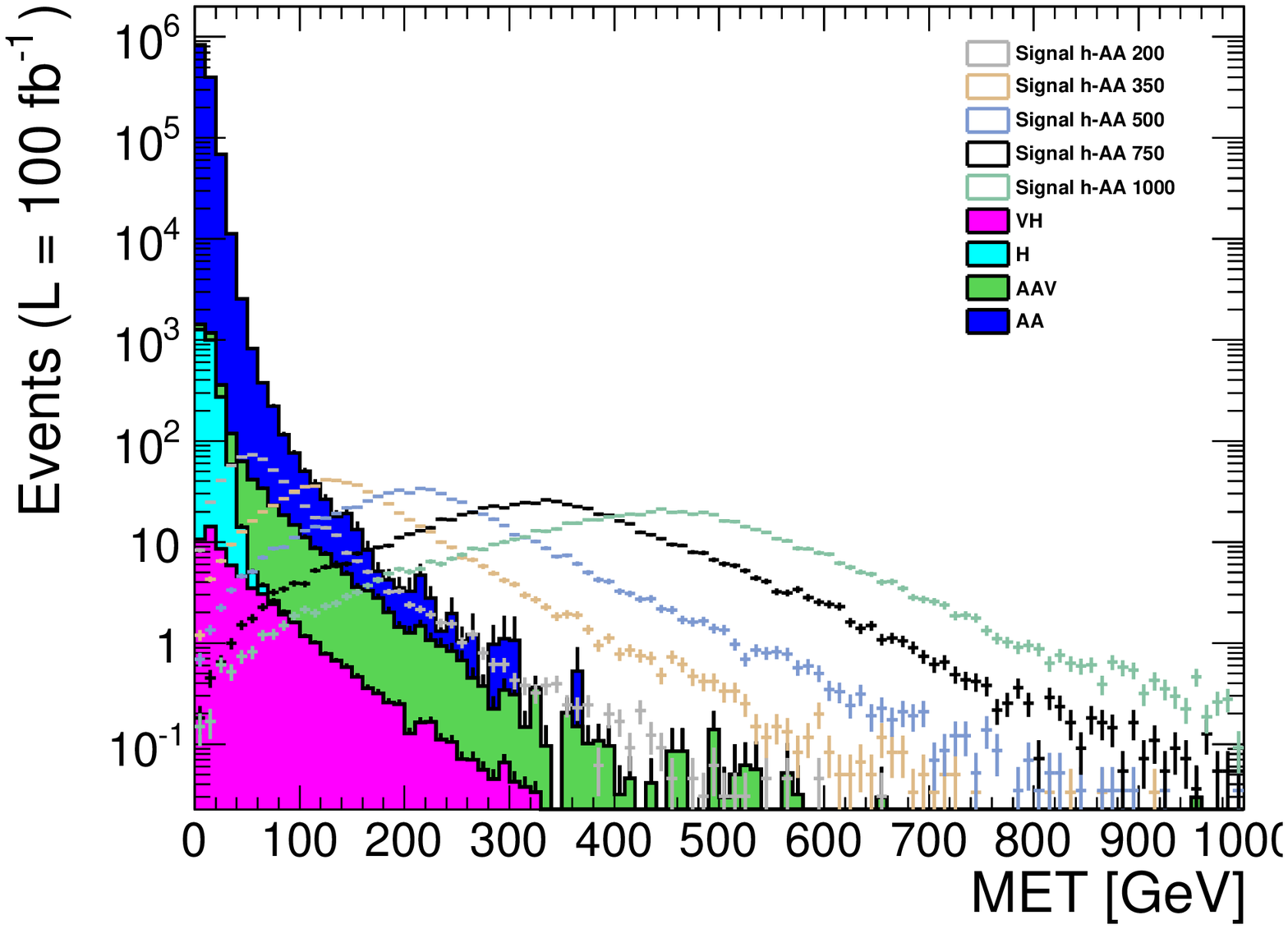}
\includegraphics[width=0.48\textwidth]{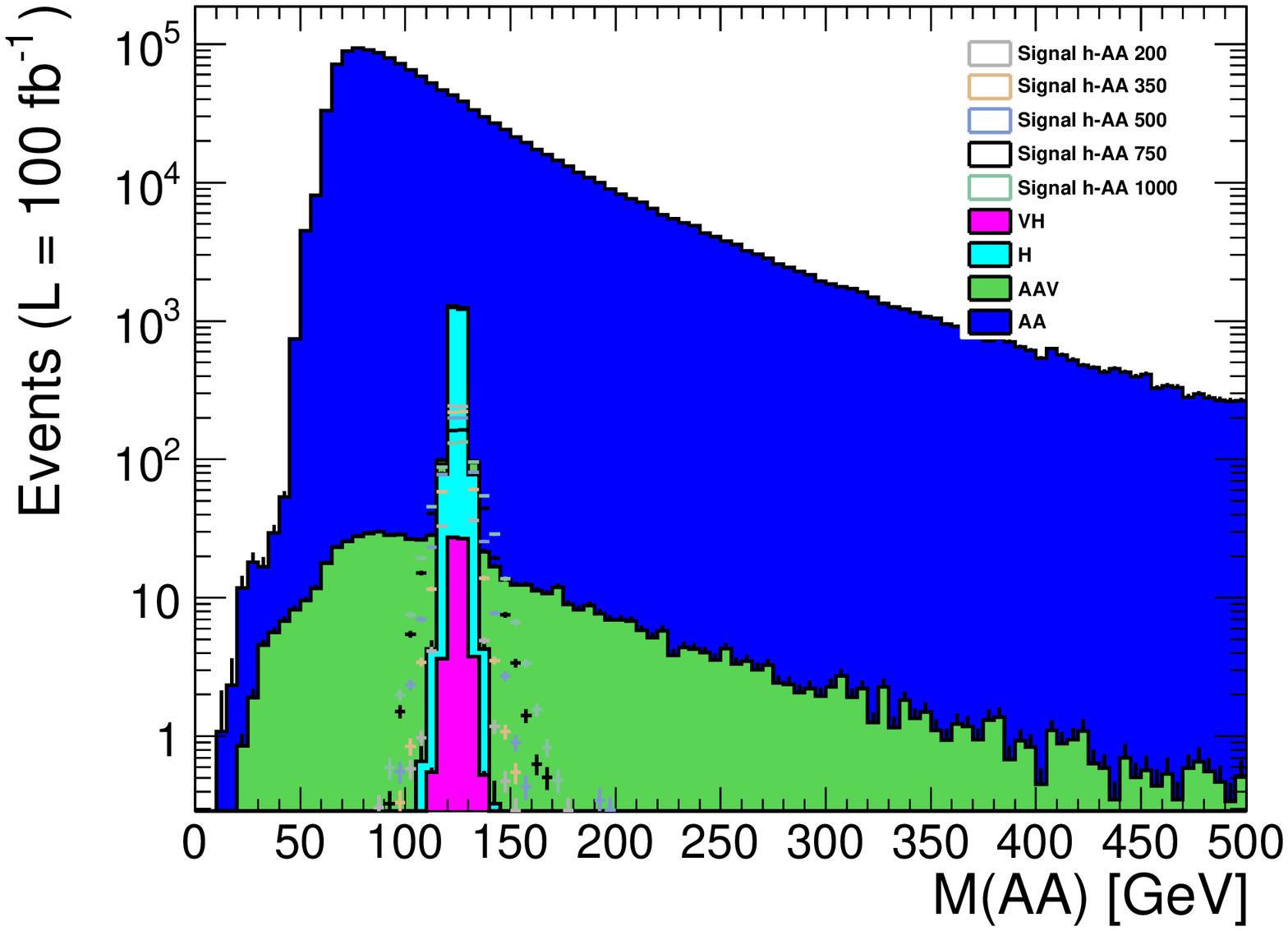}
\caption{(Left) MET, missing transverse energy. (Right) $\gamma\gamma$ invariant mass. Signal plotted for a nominal cross section of $10$~fb.\label{fig:MET-Maa}}
}

To select our Higgs $+$ MET signature, we apply the following cuts
\begin{equation}\label{eq:MET-cutaa}
{\mbox{MET}} \geq 120 {\mbox{ GeV}}\, ,
\end{equation}
\begin{equation}\label{eq:MH-cutaa}
110 \leq M(\gamma\gamma)/\mbox{GeV} \leq 140\, .
\end{equation}
These cuts have been optimised to enhance the signal-over-background ratio, compatibly with the experimental resolution. Their effect is to considerably suppress the backgrounds that have little to no MET, in particular the SM diphoton component and the SM Higgs production via gluon fusion. Eq.~(\ref{eq:MH-cutaa}) further suppresses the diphoton plus vector boson component, that has some final state MET but no Higgs. As can be seen in table~\ref{fig:PTh-MTaa}, at this stage one is left with just Higgs production processes in association with vector bosons (Higgs-strahlung) or the diphotons plus gauge bosons.

As in the previous section, the fact that the Higgs boson is produced resonantly means that it will have some transverse momentum in the case of the signal. This can be seen in figure~\ref{fig:PTh-MTaa}(left), drawn after cuts of eqs~(\ref{eq:MET-cutaa})--(\ref{eq:MH-cutaa}). The following conservative cut is then applied:
\begin{equation}\label{eq:ptH-cutaa}
p_T(\gamma\gamma) \geq 100 \mbox{ GeV}\, ,
\end{equation}
that retains at least $90\%$ of signal also for low resonant masses, while still removing some backgrounds. Finally, the resonant nature of the signal can be further exploit as in the previous section to access the intermediate resonance mass when this is larger than the Higgs mass. The less aggressive cuts here as compared to previously (in particular the MET cut of eq.~(\ref{eq:MET-cutaa}) compared to eq.~(\ref{eq:MET-cut})) enables to get a relatively sharp peak for resonances larger than just $300$ GeV, as compared to $500$ GeV in the $b$-jet final state. Suitable selections can then be further applied to single out the signal.

\FIGURE{
\centering
\includegraphics[width=0.48\textwidth]{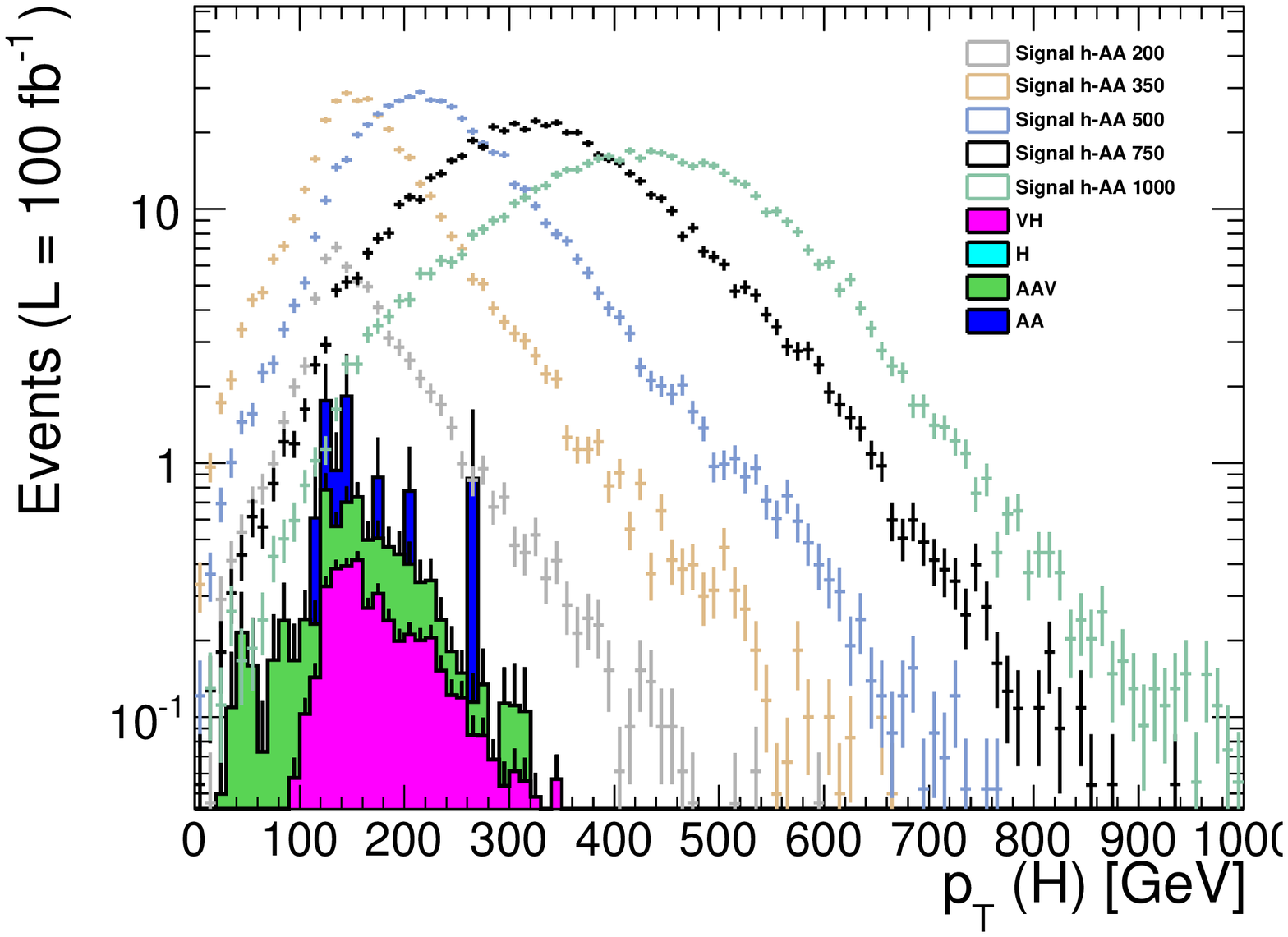}
\includegraphics[width=0.48\textwidth]{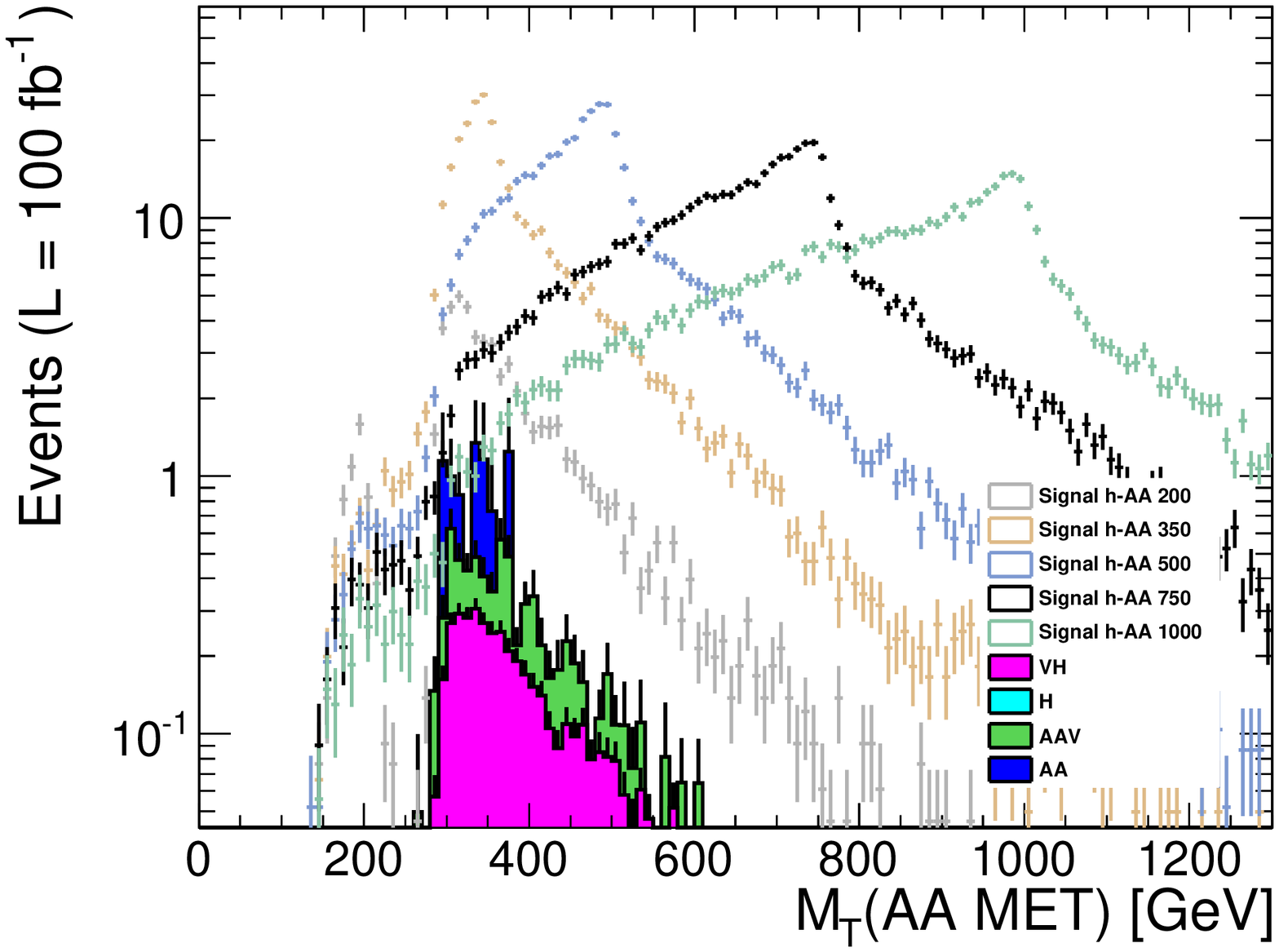}
\caption{(Left) Transverse momentum of the Higgs boson, after cut~(\ref{eq:MH-cutaa}). (Right) Transverse mass of total  Higgs $+$ MET system. Signal plotted for a nominal cross section of $10$~fb.\label{fig:PTh-MTaa}}
}

Efficiencies of the various cuts for the backgrounds and for the signal benchmarks (for a nominal normalisation cross section of $10$ fb) are collected in tables~\ref{Table:CutFlow-bkgaa}--\ref{Table:CutFlow-sgnaa}.

\begin{table}[!h]
\scalebox{0.85}{
\begin{tabular}{|l|c|c|c|c|c|c|c|c|c|}
\hline
\hline
Cut & Sum Backgrounds  & $VH$  & $H$  & $\gamma\gamma V$  & $\gamma\gamma$ \\
\hline
no cuts  & $(18658 \pm 4)\cdot 10^{3}$ & 499 $\pm$ 1 & 7431 $\pm$ 28 & 12872 $\pm$ 23 & $(18638 \pm 4)\cdot 10^{3}$ \\
$N_{\ell} \equiv 0$ & $(18658 \pm 4)\cdot 10^{3}$ & 482 $\pm$ 1 & 7431 $\pm$ 28 & 12445 $\pm$ 23 & $(18637 \pm 4)\cdot 10^{3}$ \\
$N_j \equiv 0 $ & $(13986 \pm 4)\cdot 10^{3}$ & 174 $\pm$ 1 & 4994 $\pm$ 23 & 4244 $\pm$ 13 & $(13977 \pm 4)\cdot 10^{3}$ \\
$N_\gamma \equiv 2 $ & $(1328 \pm 1)\cdot 10^{3}$ & 63 $\pm$ 0 & 2573 $\pm$ 17 & 690 $\pm$ 5 & $(1324 \pm 1)\cdot 10^{3}$ \\
MET $> 120  $ & 120 $\pm$ 7 & 5 $\pm$ 0 & 0 $\pm$ 0 & 37 $\pm$ 1 & 79 $\pm$ 7 \\
$110<M(b\overline{b})/\mbox{GeV}<140$ & 13 $\pm$ 2 & 5 $\pm$ 0 & 0 $\pm$ 0 & 4 $\pm$ 0 & 4 $\pm$ 2 \\
$p_T(H) \geq 100  $ & 12 $\pm$ 2 & 5 $\pm$ 0 & 0 $\pm$ 0 & 3 $\pm$ 0 & 4 $\pm$ 2 \\
\hline
\hline
\end{tabular}
}
\caption{ CutFlow for backgrounds for $\gamma\gamma+$MET, events for $\mathcal{L}=100$ fb$^{-1}$.}
\label{Table:CutFlow-bkgaa}
\end{table}

\begin{table}[!h]
\scalebox{0.8}{
\begin{tabular}{|l|c|c|c|c|c|}
\hline
\hline
Cut & Sum Backgrounds  & $h\ell$@200 GeV & $h\nu$@200 GeV & $h\ell$@350 GeV & $h\nu$@350 GeV \\
\hline
no cuts  & $(18658 \pm 4)\cdot 10^{3}$ & 1000 $\pm$ 4 & 1000 $\pm$ 4 & 1000 $\pm$ 4 & 1000 $\pm$ 4  \\
$N_{\ell} \equiv 0$ & $(18658 \pm 4)\cdot 10^{3}$ & 916 $\pm$ 4 & 1000 $\pm$ 4 & 956 $\pm$ 4 & 1000 $\pm$ 4 \\
$N_j \equiv 0 $ & $(13986 \pm 4)\cdot 10^{3}$ & 558 $\pm$ 3 & 700 $\pm$ 3 & 530 $\pm$ 3 & 642 $\pm$ 3 \\
$N_\gamma \equiv 2 $ & $(1328 \pm 1)\cdot 10^{3}$ & 252 $\pm$ 2 & 322 $\pm$ 2 & 265 $\pm$ 2 & 338 $\pm$ 2 \\
MET $> 120$ GeV & 120 $\pm$ 7 & 13 $\pm$ 0 & 61 $\pm$ 1 & 160 $\pm$ 2 & 221 $\pm$ 2 \\
$110<M(\gamma\gamma)/\mbox{GeV}<140$ & 13 $\pm$ 2 & 12 $\pm$ 0 & 60 $\pm$ 1 & 152 $\pm$ 2 & 216 $\pm$ 2 \\
$p_T(H) \geq 100$ GeV & 12 $\pm$ 2 & 9 $\pm$ 0 & 56 $\pm$ 1 & 118 $\pm$ 1 & 211 $\pm$ 2 \\
\hline
$ M_T (\gamma\gamma$ MET$) \geq 400$ GeV & $3.53 \pm 0.23$ & $-$ & $-$ & $-$ & $-$ \\
$ M_T (\gamma\gamma$ MET$) \geq 550$ GeV & $0.75 \pm 0.08$ & $-$ & $-$ & $-$ & $-$ \\
$ M_T (\gamma\gamma$ MET$) \geq 600$ GeV & $0.47 \pm 0.05$ & $-$ & $-$ & $-$ & $-$ \\
\hline
\end{tabular}
}
\scalebox{0.765}{
\begin{tabular}{|l|c|c|c|c|c|c|}
\hline
Cut & $h\ell$@500 GeV & $h\nu$@500 GeV & $h\ell$@750 GeV & $h\nu$@750 GeV & $h\ell$@1 TeV & $h\nu$@1 TeV \\
\hline
no cuts  & 1000 $\pm$ 4 & 1000 $\pm$ 4 & 1000 $\pm$ 4 & 1000 $\pm$ 4 & 1000 $\pm$ 4 & 1000 $\pm$ 4 \\
$N_{\ell} \equiv 0$ & 971 $\pm$ 4 & 1000 $\pm$ 4 & 979 $\pm$ 4 & 1000 $\pm$ 4 & 984 $\pm$ 4 & 1000 $\pm$ 4\\
$N_j \equiv 0 $ & 507 $\pm$ 3 & 607 $\pm$ 3 & 485 $\pm$ 3 & 573 $\pm$ 3 & 470 $\pm$ 3 & 554 $\pm$ 3 \\
$N_\gamma \equiv 2 $ &  280 $\pm$ 2 & 357 $\pm$ 2 &  290 $\pm$ 2 & 374 $\pm$ 3 & 280 $\pm$ 2 & 363 $\pm$ 3\\
MET $> 120$ GeV & 241 $\pm$ 2 & 315 $\pm$ 2 & 276 $\pm$ 2 & 360 $\pm$ 3 & 273 $\pm$ 2 & 357 $\pm$ 3 \\
$110<M(\gamma\gamma)/\mbox{GeV}<140$ & 225 $\pm$ 2 & 304 $\pm$ 2 & 243 $\pm$ 2 & 330 $\pm$ 2 & 228 $\pm$ 2 & 309 $\pm$ 2\\
$p_T(H) \geq 100$ GeV & 210 $\pm$ 2 & 302 $\pm$ 2 & 238 $\pm$ 2 & 238 $\pm$ 2 & 226 $\pm$ 2 & 309 $\pm$ 2 \\\hline
$ M_T (\gamma\gamma$ MET$) \geq 400$ GeV & $173 \pm 2$ & $248 \pm 2$ & $-$ & $-$ & $-$ & $-$ \\
$ M_T (\gamma\gamma$ MET$) \geq 550$ GeV & $-$ & $-$ & $187 \pm 2$ &$249 \pm 2$ & $-$ & $-$ \\
$ M_T (\gamma\gamma$ MET$) \geq 600$ GeV & $-$ & $-$ & $-$ & $-$ & $193 \pm 2$ & $249 \pm 2$ \\
\hline
\hline
\end{tabular}
}
\caption{ CutFlow for signals for $\gamma\gamma+$MET, events for $\mathcal{L}=100$ fb$^{-1}$ for a nominal cross section of $10$ fb.}
\label{Table:CutFlow-sgnaa}
\end{table}

The excluded cross sections evaluated as previously discussed are collected in table~\ref{tab:xsExcl-aa} and displayed in Figure~\ref{fig:xsExcl}.

\begin{table}[!h]
\centering
\scalebox{0.78}{
\begin{tabular}{|l|c|c|c|c|c|}
\hline
\hline
$h\to \gamma\gamma$ & $M=200$ GeV  & $M=350$ GeV  & $M=500$ GeV  & $M=750$ GeV  & $M=1000$ GeV \\
\hline
$\sigma$ (fb), $\mathcal{L}=100$ fb$^{-1}$ & $2.949\pm 0.010$ & $0.538\pm 0.005$ & $0.297\pm 0.002$ & $0.213\pm 0.002$ &$0.201\pm 0.002$\\
$\sigma$ (fb), $\mathcal{L}=3$ ab$^{-1}$ & $0.4306\pm 0.0020$ & $0.0488\pm 0.0004$ & $0.0358\pm 0.0003$ & $0.0179\pm 0.0001$ &$0.0148\pm 0.0001$\\
\hline
\hline
\end{tabular}
}
\caption{ Excluded cross sections in the $\gamma\gamma$ final state, at the LHC for $\sqrt{s}=13$ TeV.}
\label{tab:xsExcl-aa} 
\end{table}

\subsection{Exclusions}\label{sect:exclusions}

\begin{figure}[!h]
\centering
\includegraphics[width=0.48\textwidth]{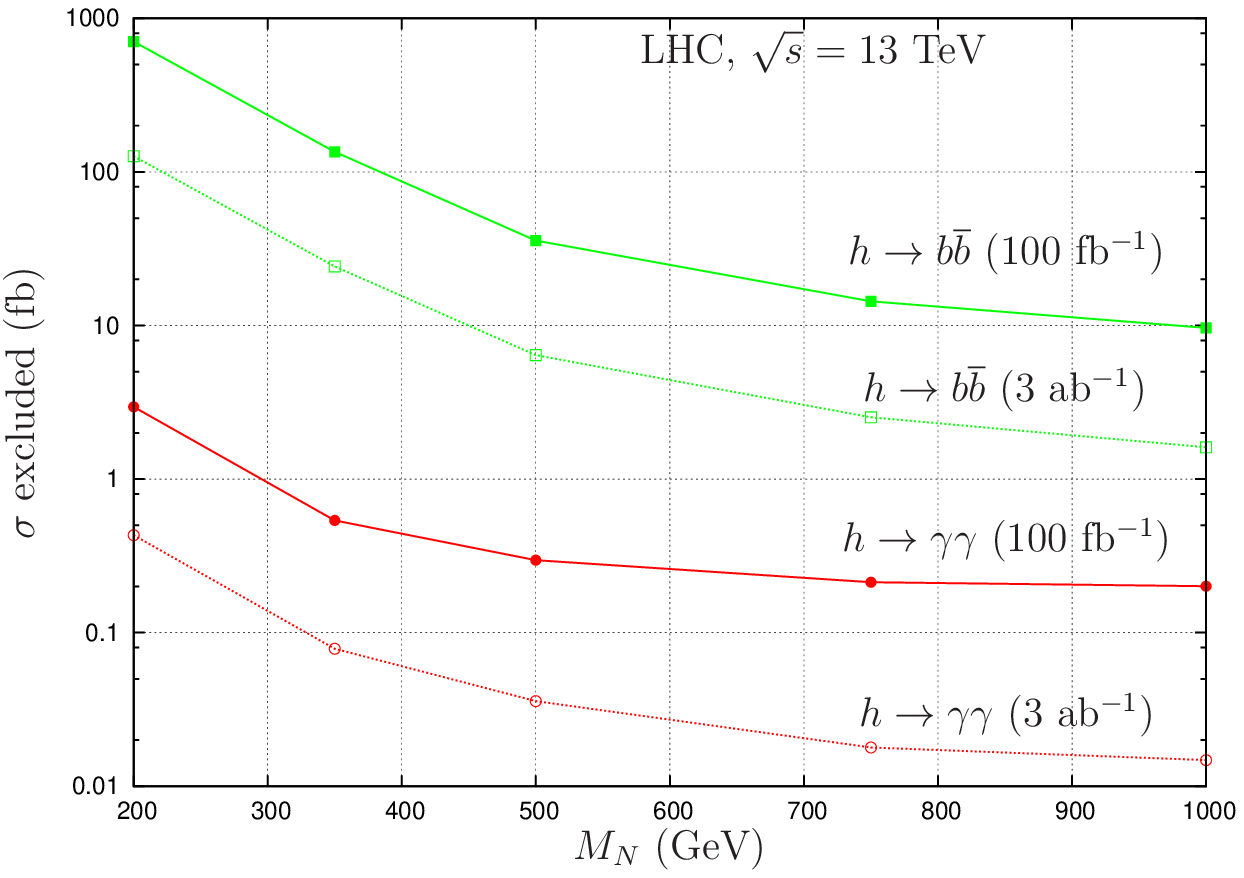}
\includegraphics[width=0.48\textwidth]{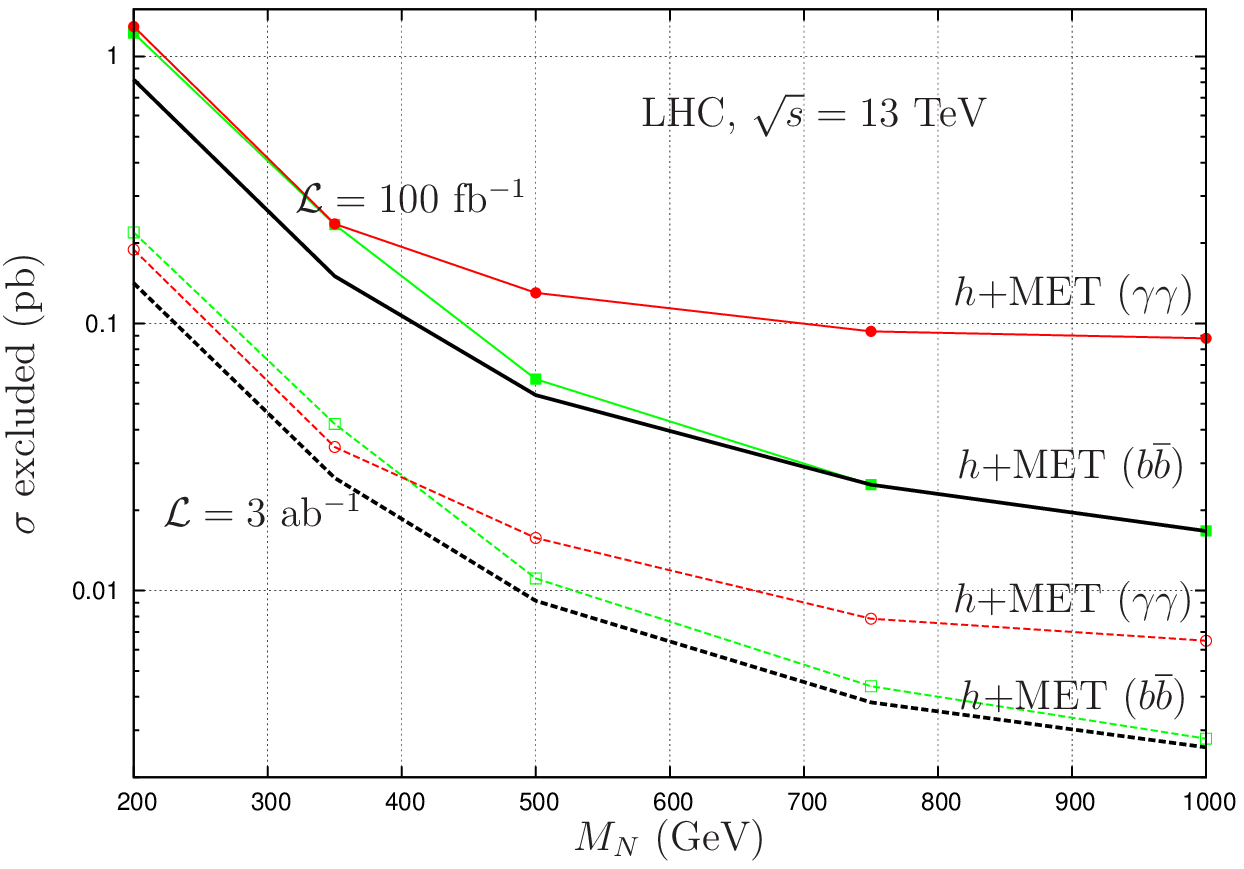}
\caption{Excluded cross sections (left) for the considered Higgs decay patterns and (right) for the Higgs boson + MET. The black lines represent the combined exclusions.\label{fig:xsExcl}}
\end{figure}

In figure~\ref{fig:xsExcl} we summarise and compare the excluded cross sections at the LHC run-II for $\sqrt{s}=13$ TeV both with $\mathcal{L}=100$ fb$^{-1}$ and $\mathcal{L}=3$ ab$^{-1}$ of data, for a resonant Higgs $+$ MET production in the Higgs-to-$b\overline{b}$ and Higgs-to-diphoton channels. Separate channels are presented in figure~\ref{fig:xsExcl}(left), while the Higgs $+$ MET production (assuming the SM Higgs branching ratios~\footnote{Although in general the presence of BSM particles might affect the Higgs branching ratios, here we assumed the SM ones for a more model-independent evaluation.}) is shown in figure~\ref{fig:xsExcl}(right). Contrary to the diphoton case, where the background after all the cuts is already negligible, the exclusion power of the $b\overline{b}$ final state steadily grows with the mass of the intermediate resonance. This is because the background rejection power of the cuts improves when the latter increases. The diphoton final state can generally exclude cross sections that are a factor $50\div 240$ lower than the $b\overline{b}$ case. However, $\displaystyle BR(h\to b\overline{b})/BR(h\to \gamma\gamma) = 0.577/(2.28\cdot 10^{-3}) \simeq 250$. Hence, a signal comprising the SM Higgs boson in the final state will get better constrained by the $b\overline{b}$ channel. To achieve the most stringent bounds, in figure~\ref{fig:xsExcl}(right) we also show the combination of the two final states. This is done employing a likelihood profiling methods for hypothesis testing with no systematic errors~\footnote{The case of no systematic errors corresponds to the formula of eq.~\ref{eq:significance}. This is done for consistencies and to draw the most optimal case. Systematic errors will rapidly degrade the results here presented, and can be quantitatively address only by a suitable experimental analysis, that goes beyond the scope of this paper especially in the light of the approximations taken through this work, i.e. the use of fast simulation.}, based on RooStats~\cite{Moneta:2010pm}.

\subsubsection{Reinterpretation in the SPSS}\label{sect:reinterpretation}
\begin{figure}[!h]
\centering
\includegraphics[width=0.75\textwidth]{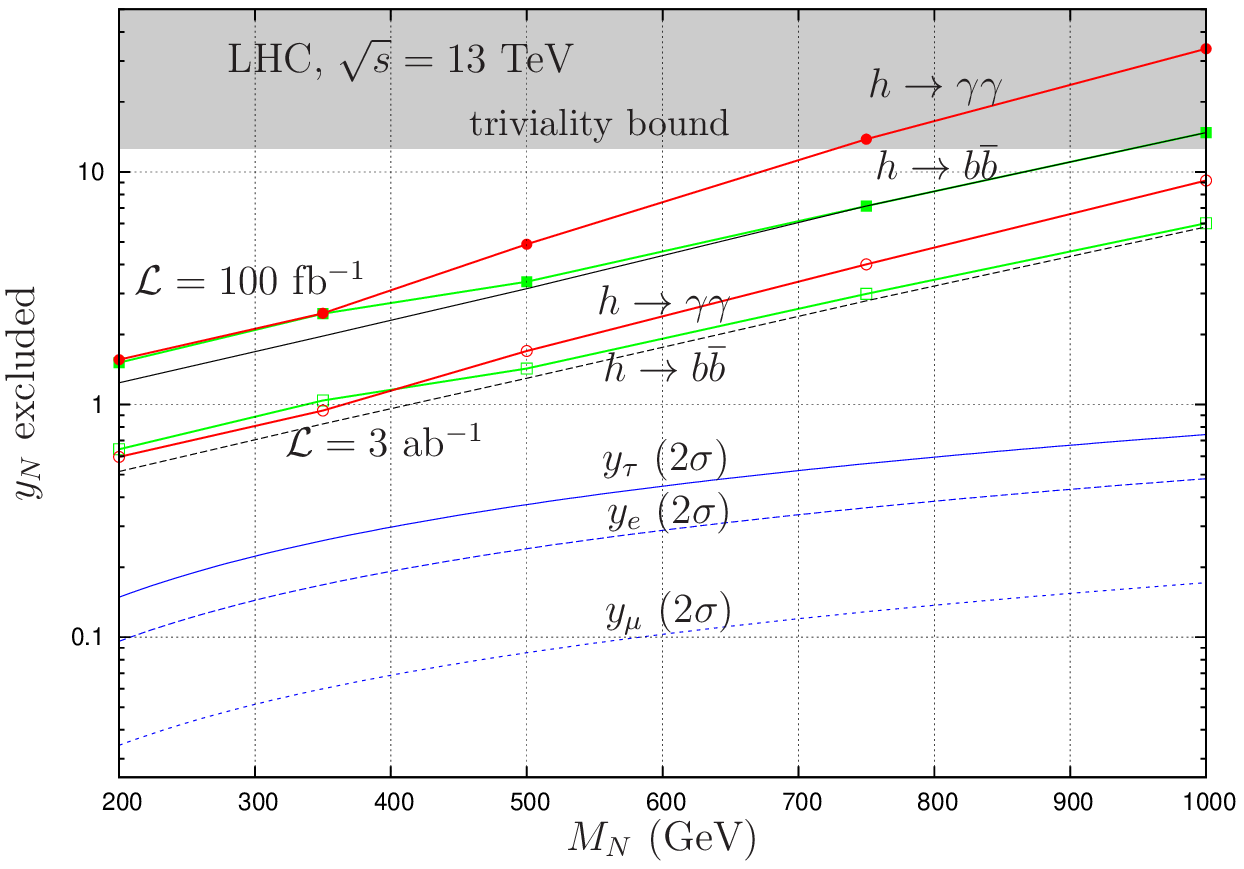}
\caption{Excluded Yukawa couplings in the SPSS for the considered Higgs decay patterns. The black lines represent the combined exclusions.\label{fig:yNExcl}}
\end{figure}

Finally, we can recast our excluded cross sections in the framework of the SPSS introduced in section~\ref{sect:model}.
Exclusions in the $y_N$--$M_N$ plane are presented in figure~\ref{fig:yNExcl}. At low $M_N$, the diphoton and $b\overline{b}$ final states have a very similar exclusion powers. However the two channels are rather different; the diphoton case is basically background free, and hence exclusions are just limited by the signal cross section to yield an observable amount of events, while for the $b\overline{b}$ case, the signal has to be at least twice the fluctuation of the large background. Therefore, the exclusion power of the latter gets better than the photons' one as the resonance mass increases because the selections that are applied also have an increasing performance at larger $M_N$. Instead, there the diphoton final state is already background free, and the cuts performance cannot increase. At the same time, the fact that the background is negligible also explains why the exclusion power of the diphoton final state improves faster than the $b\overline{b}$ one when increasing the amount of data. As before, we also show the combination of the two final states to give the most stringent bound.

In the same figure we also show the triviality bound $y_/nu < 4\pi$, that is the intrinsic upper bound to have a perturbative theory, and the present exclusion from direct and indirect constraints, see eqs.~(\ref{eq:bounde})--(\ref{eq:boundtau}). Notice that these bounds have been rescaled by a factor $2$ in the figure to corresponds to a $2\sigma$ exclusion, starting from the $68\%$ ($1\sigma$) Bayesian confidence level. We see that for 100 fb$^{−1}$ of data only neutrino masses below $900$ GeV are in the non-perturbative regime, while the exclusions at the ultimate $3$ ab$^{−1}$ are well below it. However, they are still roughly an one order of magnitude above the current constraints, here plotted separately per each flavour.

\section{Conclusions}\label{sect:conclusions}
The fact that the Higgs boson properties are going to be thoroughly investigated at the LHC in the coming years, allows one to consider rare phenomena involving the Higgs boson itself. The case of the mono Higgs signature, i.e., its production in association with just missing energy, has been recently proposed and commonly interpreted in models explaining dark matter, where the dark matter candidates are the source of missing energy.

In this work, it was shown that the mono-Higgs signature also very commonly arises from models explaining neutrino masses. In this case and differently from most dark matter models, the Higgs boson is produced resonantly, via a fermionic resonance. 

This case was then studied at the LHC by means of a detailed fast detector simulation, to first assess the LHC discovery potential for the signature of a Higgs boson produced with large MET, exploiting its resonant nature. Contrary to the dark matter case, here the final state with a pair of $b-$jets proved to be the most sensitive one, and comparable to the diphoton final state at low resonance masses. The combination of the 2 final states was therefore pursued, showing that the LHC with 100 fb$^{-1}$ can exclude cross sections down to $\mathcal{O}(10)$ fb, and down to $\mathcal{O}(1)$ fb with the ultimate 3 ab$^{-1}$ of integrated luminosity, for intermediate resonance masses up to 1 TeV.

Finally, the mono-Higgs signature has been reinterpreted in a benchmark scenario for neutrino masses description, the ``symmetry protected seesaw scenario'', where the embedded symmetry allows for naturally explaining the smallness of neutrino masses when slightly broken. For this scenario, the exclusions derived by the mono-Higgs signature are not competitive with present bounds coming from fits to low- and high-energy data. This situation could be improved by future lepton colliders, which could improve on the present bounds already in the first years of running, see Ref.~\cite{Antusch:2015gjw}.

It is worth mentioning that the mono-Higgs signature is not the one with highest sensitivity to this model. One expects the heavy neutrino decay into a lepton and 2 jets (via a W boson) to have the largest sensitivity. However, the choice of studying the mono-Higgs signature was dictated by the fact that in the Literature this is associated just to dark matter models, while it equally arises in neutrino mass models. This paper therefore is intended to offer the experimental community a new framework for the reinterpretation of upcoming results concerning the Higgs boson.

Further reinterpretions of this signature are possible in $Z'$ models, where the intermediate resonance is a vector boson. A resonant mono-Higgs signature can arise as $Z' \to h Z$, $Z\to \nu \nu$ in general models where the $Z'$ boson mixes with the $Z$ boson, or from heavy neutrino pair production in a gauged type-I/inverse seesaw model, such as in the $B-L$ model, when one neutrino decays into $\nu h$ and the other one into $\nu Z$, $Z\to \nu \nu$~\cite{Basso:2008iv,Basso:2012ti,Kang:2015uoc}. Due to the focus of this work on fermionic-mediated signatures, the above $Z'$-mediated ones have not been discussed, and will be analysed in the future.

\section*{Acknowledgements}
I thank Oliver Fischer, Stefan Antusch and Eros Cazzato for valuable discussions at the beginning of this work and for reading the manuscript.
This work has been partially supported by the French ANR 12 JS05 002 01 BATS@LHC and by the Theory-LHC-France initiative of the CNRS/IN2P3,
and partially by the OCEVU Labex (ANR-11-LABX- 0060) and the A*MIDEX project (ANR-11-IDEX-0001-02), funded by the “Investissements d’Avenir” French government program managed by the ANR.

%\clearpage
\bibliographystyle{h-physrev5}
\bibliography{monoH}

\end{document}